\documentclass[useAMS,usenatbib]{mn2e}
\usepackage{amsmath}%\usepackage{natbib}
\usepackage{nccmath} %Fleqn
\usepackage{stfloats}
\usepackage{epsfig}
\usepackage{natbib}
\usepackage{subfigure}
\usepackage{multirow}
\usepackage{amssymb}
\usepackage{gensymb} 
\usepackage{color}
\def \Mdot 	{$ \text{M}_\odot $}
\def \Ldot 		{ \text{L}_\odot } 
\def \Ldotpc 	{\Ldot \ pc^{-2}}
\def \lbox 		{17cm}
\def \lfull 		{12cm}
\newcommand{\Cint}{C_{\rm int}}
\newcommand{\Crad}{C_{\rm rad}}
\newcommand{\Csig}{C_{\sigma}}

\def \ngal {4342} %Number of galaxies in SAM
\def \SFit  						{\log \frac{\ I_{\mathrm{e}}}{\mathrm{\Ldotpc}} = -1.59\ \mathrm{log}\frac{R_{\mathrm{e}}}{\mathrm{kpc}} + 1.65\ \mathrm{log}\ \frac{\sigma}{\mathrm{km \ s^{-1}}} -0.49}	
\def \S12it 					{\log \frac{\ I_{\mathrm{e}}}{\mathrm{\Ldotpc}} = -1.21\ \mathrm{log}\frac{R_{\mathrm{e}}}{\mathrm{kpc}} + 1.16\ \mathrm{log}\ \frac{\sigma}{\mathrm{km \ s^{-1}}} +0.55}	%Linear FP fit equation
\def \SFitPercent					{90.2\%}										%Percentage of galaxies in middle 3 FP slices
\def \SpearmanAgeSig {0.74}
\def \SpearmanAgeRe {0.03}
\def \SpearmanZSig {0.83}
\def \SpearmanZRe {0.36}
\title{Modeling the Ages and Metallicities of Early-Type Galaxies in Fundamental Plane Space}
\author[Porter et al.]{L.~A. Porter$^{1,2}$, R.~S. Somerville$^{3}\footnotemark[1]$, J.~R. Primack $^{1,2}$, D.~J. Croton$^{4}$, \newauthor M.~D. Covington$^{1,2,5}$, G.~J. Graves$^{6}$ and S.~M. Faber$^{7}$ \\
     $^1$Department of Physics, University of California, Santa Cruz, California 95064, USA\\
	$^2$Santa Cruz Institute for Particle Physics, University of California, Santa Cruz, California 95064, USA\\	
     $^3$Department of Physics and Astronomy, Rutgers University, Piscataway, New Jersey 08854, USA\\
     $^4$Centre for Astrophysics and Supercomputing, Swinburne University of Technology, Melbourne, Australia\\
     	$^5$NSF International Research Fellow, Karst Research Institute ZRC SAZU, Titov trg 2, SI-6230 Postojna, Slovenia\\
	$^6$Department of Astrophysical Sciences, Peyton Hall, Princeton, NJ 08540\\
     $^7$ UCO/Lick Observatory, Department of Astronomy and Astrophysics, University of California, Santa Cruz, CA 95064, USA\\
     }
	
\begin{document}
\maketitle
\begin{abstract}
Recent observations have probed the formation histories of nearby
elliptical galaxies by tracking correlations between the stellar
population parameters, age and metallicity, and the structural
parameters that enter the Fundamental Plane, size $\mathrm{R_{e}}$
and velocity dispersion $\sigma$.   {\color{black} These studies have found
intriguing correlations between these four parameters.} 
In this work, we make use of a semi-analytic model,
based on halo merger trees extracted from the Bolshoi cosmological
simulation, that predicts the structural properties of
spheroid-dominated galaxies based on an analytic model that has been
tested and calibrated against an extensive suite of
hydrodynamic+N-body binary merger simulations.  We predict the
$\mathrm{R_{e}}$, $\sigma$, luminosity, age, and
metallicity of spheroid-dominated galaxies, enabling us to compare
directly to observations.
Our model predicts a strong correlation between age and $\sigma$ 
for early-type galaxies, and no significant correlation
between age and radius, in agreement with observations. In addition we
predict a strong correlation between metallicity and $\sigma$, 
and a weak correlation between metallicity and
$\mathrm{R_{e}}$, in qualitative agreement with observations. 
%This is also in qualitative agreement with observations,
%although our predicted dependence on radius is stronger than some
%observational studies have found.
%
{\color{black} We find that the correlations with $\sigma$ arise as a
result of the strong link between $\sigma$ and the galaxy's assembly time.}  
Minor mergers produce a large change in
radius while leaving $\sigma$ nearly the same, which
explains the weaker trends with radius.
\end{abstract}
	\footnotetext[1]{email: somerville@physics.rutgers.edu}
\section{Introduction}

Nearby galaxies are commonly divided into two basic morphological
types: spheroid-dominated, ``early type'' galaxies, and
disk-dominated, ``late type'' galaxies. Early type galaxies are
dominated by random motions, have compact, concentrated light
profiles, and are typically red and gas poor, while late-type galaxies
are rotation supported, have more extended light profiles, and tend to
be gas-rich, blue, and star forming.

Early type galaxies lie on a two-dimensional plane relating effective
radius ($R_\mathrm{{e}}$), central stellar velocity dispersion
($\sigma$), and effective surface brightness ($I_{\mathrm{e}}$),
termed the fundamental plane (FP)
\citep{Djorgovski:1987b,Dressler:1987a,Faber:1987a}.  This plane is
tilted from the plane one would expect from a simple application of
the virial theorem, indicating that further processes, such as
non-homology  {\color{black}(i.e., differences in the density profile and orbital 
distribution)} or a varying mass-to-light ratio, must have an effect
\citep{Jorgensen:1996b}.  Furthermore, the fundamental plane is not an
exact relation; galaxies have a degree of scatter around the FP, in
effect making the fundamental plane `thick'. Observations indicate
that this scatter increases with redshift, particularly among less
massive galaxies \citep{Treu:2005b}.

More specifically, while the slope of the FP appears unchanged for
high-mass ellipticals since $z \sim 1$, low-mass ellipticals at high
redshifts have higher surface brightnesses than their effective radii
and velocity dispersions would seem to predict
\citep{Wel:2004a,Treu:2005a,Treu:2005b,Jorgensen:2006a,Dokkum:2007a}.
If we consider a projected FP, where surface brightness is the
dependent parameter, then these low-mass galaxies tend to lie above
the mean FP relation {\color{black} at high redshift} (i.e., they are brighter 
than average), and to fall onto the FP over time.

There are indications that this residual thickness in the FP
correlates with the stellar population age.  \cite{Forbes:1998a} and
\cite{Terlevich:2002a} found that galaxies with higher residual
surface brightnesses are younger than those that lie near the
mid-plane of the FP; conversely, those with lower residual surface
brightnesses are older.

Recently, observations from the Sloan Digital Sky Survey (SDSS) have
been used to analyze stellar population trends both within the
R-$\sigma$ projection of the FP, and through the thickness of the FP,
using residual surface brightnesses
\citep{Graves:2009c,Graves:2009b,Graves:2010b,Graves:2010a}.  By
stacking spectra of galaxies with similar stellar properties and
measuring the Lick indices on those spectra, the authors were able to
derive [Fe/H],[Mg/H], [Mg/Fe], and stellar age for a population of
passive early-type galaxies. In agreement with \cite{Forbes:1998a} and
\cite{Terlevich:2002a}, Graves et al.~(2009b, hereafter G09) found
that younger galaxies lie above the FP, and have relatively higher
surface brightnesses, while older galaxies lie below it. Galaxies
above the FP also tended to have higher [Fe/H] and [Mg/H], and lower
[Mg/Fe].  G09 also determined that age, [Fe/H], [Mg/H], and [Mg/Fe]
increase with velocity dispersion throughout the FP, independent of
the residual surface brightness.  These same properties are almost
independent of $R_\mathrm{{e}}$, {\color{black} indicating that a galaxy's velocity
dispersion, and not its dynamical mass ($\propto \sigma^{2}R_{e}$),
has the better correlation with its star formation history.}
The strong dependence of age and metallicity on velocity dispersion is
consistent with previous studies \citep{Smith:2007a,Nelan:2005a} {\color{black}
and more recent studies \citep{Thomas:2010,Greene:2012,Johansson:2012b}.}
Similar results were obtained in a recent analysis of the Six-degree
Field Galaxy Survey (6dFGS) \citep{Jones:2004a,Jones:2009a} by
\cite{Springob:2012a}, though this work finds a slightly stronger
dependence of age and metallicity on effective radius.  

The physical origin of the Fundamental Plane, its slope and evolution,
and correlations between the structural parameters that enter it and
stellar population parameters such as age and metallicity remain
somewhat poorly understood. A wide variety of mechanisms have been
proposed to explain the formation of spheroids and the apparent
transformation of gas rich, star forming, disk-dominated galaxies into
gas poor, passive spheroid-dominated galaxies that seems to be implied
by observations of high redshift galaxies \citep{bell:07,faber:07}.
Based on an early generation of numerical simulations, a ``merger''
picture was proposed in which elliptical galaxies are formed through
``major'' (near equal-mass) mergers of disk galaxies
\citep[e.g.][]{Toomre:1972a,Toomre:1977a,Mihos:1994a,Barnes:1996b}. More
recently, it has been realized that the picture is more probably more
complex: the chemical, dynamical, and structural properties of local
giant ellipticals are not consistent with having been formed through a
single binary merger between progenitors similar to nearby large
spirals like the Milky Way. Rather, mergers at high redshift probably
involve progenitors that are denser and more gas rich than our Galaxy,
leading to more compact remnants
\citep{Dekel:2006a,Ciotti:2007a,Wuyts:2010b,Khochfar:2003a}.
Subsequent dissipationless `dry' mergers and minor mergers can greatly
increase the galaxies' stellar masses and radii  {\color{black}
\citep{Naab:2006b,Naab:2009a,Bezanson:2009a,Oser:2012a,Laporte:2013,Hilz:2013}.} 
Thus the massive ellipticals
seen in the local universe are probably built up through a complex
sequence of multiple mergers, including ``wet'' (gas-rich), moist and
dry mergers, major and minor mergers, and mergers between progenitors
with a variety of morphologies and sizes \citep{khochfar:06}.

However, mergers may not be the only possible way to create a
spheroid-dominated galaxy. It has been proposed that spheroids may
also form through in-situ processes associated with gravitational
instabilities, such as via the formation of a bar that destabilizes
the disk, transferring mass into a spheroid component
\citep{Toomre:1964a,hohl:71,OP:73,combes:90,debattista:04} and via
clumps of gas that form in the disk and migrate inwards
\citep{Dekel:2009a,Bournaud:2011a,Dekel:2013a}.  Some studies based on
semi-analytic models suggest that formation of spheroids through a
non-merger channel is necessary in order to account for the observed
fraction of spheroid-dominated galaxies at intermediate masses
\citep{parry:09,De-Lucia:2011a,Porter:2014a}.  However, the efficiency
of spheroid formation via disk instabilities, its physical basis, and its
importance relative to mergers remain poorly understood issues.

%rss orphaned text -- move somewhere
%G09 posited that the lack of dependence on $R_{e}$ seen in their data
%occurs because effective radius is strongly dependent on the orbital
%parameters of the major merger and subsequent dissipationless mergers.
%This dependence introduces a large amount of scatter in the remnants'
%radii, effectively diluting any corresponding radial relations that
%were present in the progenitors.

Currently available numerical hydrodynamic simulations of cosmological
volumes typically do not have adequate resolution to robustly resolve
the internal structure of galaxies. Moreover, it is well-known that
the observed properties of early type galaxies cannot be reproduced
without introducing a process that quenches star formation and
prevents overcooling in massive objects, such as feedback from Active
Galactic Nuclei (AGN). Implementing black hole growth and AGN feedback
in a physical manner requires even higher resolution. Therefore,
although much recent progress has been made, making accurate
predictions of galaxy internal structure for statistically robust
samples remains challenging for numerical simulations.
%This remains challenging for numerical
%hydrodynamical simulations, although much recent progress has been
%made. 
%rss add some refs?

Semi-analytic models provide an alternative method to simulate the
formation and evolution of galaxies in a cosmological context. 
{\color{black} SAMs are also unable to resolve the internal structure of galaxies. The 
advantage instead is that alternative prescriptions for physical processes 
can be more efficiently implemented into SAMs, and a larger sample of 
model galaxies can be studied.}
Recent
SAMs that include schematic recipes for AGN feedback have proven to be
quite successful in reproducing a variety of observed global galaxy
properties
\citep[e.g.][]{Croton:2006a,Bower:2006a,Somerville:2008a,Fontanot:2009a,Guo:2011a,Somerville:2012a}.
In \citet[][P14]{Porter:2014a}, we incorporated a model for predicting
the sizes and velocity dispersions of spheroids formed in mergers and
disk instabilities within the ``Santa Cruz'' semi-analytic model of
Somerville et al. \citep{Somerville:2008a,Somerville:2012a}. In this
new generation of models, the SAM is implemented within merger trees
extracted from the Bolshoi cosmological N-body simulation
\citep{Klypin:2011a,Trujillo-Gomez:2011a} 
{\color{black} using the ROCKSTAR halo finder 
and the gravitationally consistent merger tree algorithm
developed by \citet{Behroozi:2013a,behroozi:13}.} The model for the structural
properties of merger remnants is motivated and calibrated based on
high-resolution numerical Smoothed Particle Hydrodynamic (SPH)
simulations of binary galaxy mergers
\citep{Cox:2004b,Cox:2006a,Cox:2008b,Johansson:2009a}. An earlier
version of this model was presented in \citet[][C08]{Covington:2008b}, and
applied in post-processing to mergers extracted from SAMs in
\citet[][C11]{Covington:2011a}. 

In P14, we extended the model of C08 to treat mergers between
progenitors spanning a wider variety of gas fractions and morphology,
including gas-poor and spheroid-dominated galaxies, using the
simulation suite of \citet{Johansson:2009a}. P14 implemented this more
general model for the structural properties of merger remnants, along
with a simple model for estimating the structural properties of
spheroids formed via disk instabilities, self-consistently within the
Santa Cruz SAM, and showed that the new SAM reproduces the observed
stellar mass function and Fundamental Plane scaling relations of
spheroid-dominated galaxies in the local universe. In addition, the
model qualitatively reproduces the observed evolution of the size-mass
relation for spheroid-dominated galaxies from $z\sim 2$--0, and the
steeper slope, smaller scatter, and more rapid evolution of the
size-mass relation for spheroid-dominated relative to disk-dominated
galaxies. The model also predicts weaker evolution in the
Faber-Jackson (mass-velocity dispersion) relation than in the
size-mass relation, in agreement with observations.

A key aspect of our model for spheroid structure is the accounting for
dissipation during mergers of progenitors that contain significant
amounts of gas. Unlike stars, gas can radiate energy away, and
therefore mergers between gas-rich progenitors result in more compact
remnants \citep{Covington:2008b,Hopkins:2008a}. Previous SAMs that
attempted to model spheroid sizes without accounting for this
dissipation were not able to reproduce observed structural scaling
relations \citep{Cole:2000a,Shankar:2011a,Guo:2011a}.

In this paper, we use the model developed by P14 to make predictions
for the relationship between early-type galaxies' \emph{structural
  parameters} (radius and velocity dispersion) and their \emph{stellar
  population parameters} (age and metallicity). We select
spheroid-dominated galaxies from our models and determine their
location within the FP using the same approach as that of the
observational study of G09, to which we compare our predictions
explicitly. We then examine the predicted correlations between
galaxies' age and metallicity as a function of velocity dispersion and
radius, for slices taken below, within, and above the FP. In this way,
we hope to better understand the origin of the observed correlations.

Section~\ref{ssec:SAMmodel} briefly describes the SAM used in our
analysis.  Section~\ref{ssec:DissipModel} provides a brief overview of
the analytic model used to calculate the radius and velocity
dispersion for spheroids.  Section 3 presents a summary of the
observations of G09, to which we make direct comparisons.
We present results beginning in Section 4, in which we examine the
relationships between either age or metallicity as a function of
velocity dispersion and radius, for different slices of the early-type
population from the P14 SAM taken parallel to the FP as described
above.  We discuss the interpretation of our results in Section 5.

\section{Methods}

We provide a very brief overview of the semi-analytic model used in
this work. For more details, see
\citet{Somerville:2008a,Somerville:2012a} and P14. We also give a
brief summary of our prescription for computing the effective radii
and velocity dispersions of spheroids, as developed in
\citet{Covington:2008b,Covington:2011a}; and P14.

\subsection {The semi-analytic model}

\label{ssec:SAMmodel}

The P14 SAM is implemented within merger trees extracted from the
Bolshoi N-body dark matter simulation
\citep{Klypin:2011a,Trujillo-Gomez:2011a} using the ROCKSTAR algorithm
\citep{behroozi:13} and the gravitationally consistent merger trees by
\citet{Behroozi:2013a}. We follow the merging and tidal
destruction of satellites within virialized halos using a
semi-analytic model described in S08. To predict the observable
properties of galaxies, we adopt fairly standard prescriptions for
photoionization, radiative cooling, star formation, supernova
feedback, chemical evolution, and black hole growth and feedback (see
S08 and P14 for details). We model the Spectral Energy Distribution
(SED) of galaxies by convolving our predicted star formation and
chemical enrichment histories with the stellar population synthesis
models of \cite{Bruzual:2003b}. We include dust extinction using
analytic prescriptions as described in S12.

We adopt the same cosmological parameters used in the Bolshoi
simulation: $\Omega_m$ = 0.27, $\Omega_\Lambda$ = 0.73, $\emph{h}$ =
0.70, $\sigma_8$ = 0.82. These are consistent with the Wilkinson
Microwave Anisotropy Probe (WMAP) five- and seven-year results
\citep{Komatsu:2009a,Komatsu:2011a}. Throughout this work we adopt a
Chabrier \citep{Chabrier:2003a} stellar Initial Mass Function (IMF).

\subsection{Model for spheroid structural parameters}
\label{ssec:DissipModel}

To compute the structural properties of galactic spheroids, we have
built upon the model developed by C08 and C11. We first consider
spheroids that are formed in mergers. In the case of a merger without
dissipation, simple conservation of energy arguments would predict
that the internal energy of the two progenitors is conserved during
the merger:
\begin{equation}
E_{\rm init} = E_{\rm f} = 
		\Cint\sum_{i=1}^{2}G\frac{(M_{*,i}+M_{\rm new*,i})^{2}}{R_{*,i}}=\Cint G\frac{M_{*, \rm f}^{2}}{R_{\rm *, f}},
		\label{eqn:Energy}
\end{equation}
where $M_{*,i}$ is the stellar mass of each of the two progenitors,
$R_{*,i}$ are the three dimensional effective radii of the
progenitors, $M_{\rm *, f}$ and $R_{\rm *, f}$ are the stellar mass
and 3D effective radius of the merger remnant, and $\Cint$ is a
dimensionless constant relating the internal energy of the galaxy to
$GM^{2}/R$.  
{\color{black} The mass of stars formed during the merger is given by
$M_{\rm new*,i} = C_{\rm new} M_{\rm gas,i} f_k$, where $C_{\rm new}
\sim 0.3$ is obtained by fitting to the \citet{Cox:2004b} simulations, 
and $f_k \equiv \Delta E/K_{\rm tot}$ where $K_{\rm tot}$ is the total
kinetic energy of the galaxy and $\Delta E$ is the impulse between the
two galaxies (see C08 equation 3 and Appendix A).}

However, in the presence of gas, mergers can be highly
dissipative; thus the conservation of energy relation must be modified
with a term incorporating radiative losses.
Motivated by the results of hydrodynamical simulations,
C08 provided a simple parameterization of this
radiative energy loss:
\begin{equation}
		\label{eqn:Energy_radiated}
		E_{\mathrm{rad}}=\Crad\sum_{i=1}^{2}K_{i}f_{\mathrm{g},i}f_{\mathrm{k},i}(1+f_{\mathrm{k},i}),
	\end{equation}
where $K_{\emph{i}}$, $f_{\mathrm{g},i}$, and $f_{\mathrm{k},i}$ are
the total kinetic energy, baryonic gas fraction, and fractional
impulse of progenitor \emph{i}, $\Crad$ is a dimensionless constant
and the sum is over the two progenitors. The energy equation above
must be modified by including this term: $E_{\rm init} + E_{\rm
  rad} = E_{\rm f}$. We use this equation to solve for the
effective radius of the stars in the remnant, $R_{*, f}$.

The model presented in C08 and C11 was limited in that it was only calibrated
against simulations of fairly gas-rich mergers of disk-dominated
progenitors. In P14 we extended the model by calibrating it with an
additional 68 hydrodynamical simulations of binary mergers, described
in \cite{Johansson:2009a}, including both major and minor mergers of
mixed-morphology and spheroid-spheroid mergers (i.e., mergers in which
one or both progenitors contain a significant spheroidal
component). Importantly, we found that the parameters $\Cint$ and
$\Crad$ differ significantly depending on the mass ratio of the merger
and the morphology and gas content of the progenitors. A complete
table of values for these parameters, measured from the binary merger
simulations, is given in P14. The value of $f_{\rm
  rad}\equiv\Crad/\Cint$ can be thought of as characterizing the
relative importance of dissipation; high values indicate more
dissipation.  We find that this value is highest for major mergers of
two disk-dominated galaxies ($f_{\rm rad} = 5.0$), is lower for minor
mergers between two disk-dominated galaxies ($f_{\rm rad} = 2.7$) and
is zero for mergers where one or both of the galaxies is
spheroid-dominated.  This latter subset of mergers is thus essentially
dissipationless.
 
We use the virial theorem to determine the line-of-sight velocity
dispersion of the spheroid:
	\begin{equation}
		\label{eqn:velocity_dispersion} 
		\sigma^2=\left(\frac{\Csig G}{2R_{\rm f}}\frac{M_{\rm *,f}}{(1-f_{\rm{dm,f}})}\right),
	\end{equation} 
where $M_{\mathrm{*,f}}$ is the stellar mass of the spheroid,
$R_{\mathrm{f}}$ is the stellar half-mass radius of the spheroid, and
$\Csig$ is a dimensionless constant that accounts for the conversion
between the three-dimensional effective radius and the line-of-sight
projection of the velocity dispersion. We define $M_{\rm dm}$ to be
the mass of dark matter within $R_{\rm f}$, and $f_{\mathrm{dm,f}} =
M_{\rm dm}/(0.5M_{\rm *,f}+M_{\rm dm})$ to be the central dark matter
fraction of the remnant. {\color{black} (The factor of 0.5 multiplies $M_{\rm *,f}$
because $R_{\rm f}$ is the stellar half-mass radius.)}
The value of $\Csig$ is again measured from
the binary merger simulations and is found to be nearly the same in
all cases (see P14).

%rss_rv changed the order of the paragraphs

In P14 we found that when we accounted for spheroid growth through
mergers alone, our model did not produce enough intermediate mass
spheroid-dominated galaxies in the local universe. We therefore
considered models in which spheroids could form and grow via disk
instabilities. When a disk is deemed unstable according to a
Toomre-like criterion \citep{Toomre:1964a,Efstathiou:1982a}, we
transfer stars or stars and cold gas from the disk to the spheroid
component until the disk becomes marginally stable again. Following
the prescription of \cite{Guo:2011a}, we assumed that the stellar mass
transferred forms a spheroid with half-mass radius equivalent to that
of the unstable disk material, which then merges dissipationally with
any existing spheroid (i.e. $C_{\mathrm{rad}} = 0$). We presented two
models of disk instability, one in which only the stellar disk
participates in the instability, and one in which both the stars and
gas in the disk participate.  These two models were tuned separately
to reproduce the early-type stellar mass function.  In this paper we
use the `Stars+Gas DI' model described in P14. We showed in P14 that
the `Stars DI' model produces very similar results.

%rss_rv
{\color{black} We note that we do not include environmental processes
  that could lead to morphological transformation, such as galaxy
  harassment, or tidal or ram pressure stripping. However these
  processes are expected to be most important in galaxy clusters and
  should be sub-dominant in field galaxy samples, on which we focus
  here. Some readers might be concerned that neglecting stripping
  processes could affect the cold gas fractions of satellites, which
  would impact the degree of dissipation experienced in a
  merger. However, tidal stripping should not change the gas
  \emph{fraction}, and strong evidence of ram pressure stripping is
  again limited to galaxy clusters. }

\section{Summary of Observations}

We compare our findings to a recent survey of early-type galaxies from
the Sloan Digital Sky Survey (SDSS) \citep{York:2000a} Spectroscopic
Main Galaxy Survey \citep{Strauss:2002a} Data Release 6
\citep{Adelman-McCarthy:2008a}.  The sample of galaxies is described
in \cite{Graves:2009c,Graves:2009b}.  The galaxies selected were
observed in the redshift range $0.04 < z < 0.08$, with light profiles
that were both centrally concentrated and fit a de Vaucouleurs
profile.  To prevent a small proportion of young stars from biasing
the measured luminosity, G09 excluded actively star-forming galaxies.
Using colors and emission-line intensities, G09 also rejected Seyfert
hosts, low ionization nuclear emission-line region (LINER) hosts, and
transition objects, as they can host active galactic nuclei (AGN)
which have been found to have light profiles intermediate between
early- and late-types \citep{Kauffmann:2003a}.  

Using the Lick indices \citep{Worthey:1994a,Worthey:1997a} on 16,000
stacked spectra, G09 calculated mean luminosity-weighted ages and
metallicities in bins with residual surface brightness above, within,
and below the fundamental plane.  The bins covered the approximate
range $\mathrm{(0.0 < \log(\emph{R}_{e}/ kpc) <0.7\ )}$,
$\mathrm{(1.9\ < \log (\sigma/km \ s^{-1}) <2.4)}$, and $-0.3 <\Delta
\log (\emph{I}_{e}/\mathrm{\Ldotpc}) < 0.3$, where
$\Delta\ \mathrm{log}\ I_{\mathrm{e}}$ is the residual surface
brightness resulting from a log fit in radius and velocity dispersion
\citep{Graves:2010b},
\begin{fleqn}
 	\begin{equation}
		\label{eqn:Graves_FP_params} 
\S12it
	\end{equation}
 \end{fleqn}
We note that single stellar population ages derived from Lick indices,
as in G09, have been shown to be biased towards the most recent
episode of star formation, resulting in age estimates that are
systematically lower than the `true' luminosity-weighted age
\citep{Trager:2009a}.

G09 formed contours relating the mean light-weighted age and
light-weighted metallicities, [Fe/H], [Mg/H], and [Mg/Fe], to
effective radius and velocity dispersion across three slices of the
face-on projection fundamental plane.  While a different version of
the S08/S12 SAM does include a more sophisticated chemical evolution
model that discards the instantaneous recycling approximation and
tracks contributions from different elements \citep{Arrigoni:2010a},
in this work we use a simplified  {\color{black} instantaneous recycling}
chemical evolution model that tracks
only the total metallicity $Z$ and does not include the contribution
from Type I supernovae. {\color{black} (See Section 2.5 of P14 for details.)} 
We thus consider the SAM metallicity to be
most similar to [Mg/H], a measure of $\alpha$-type enrichment.

The relevant results can be seen in Figures 7 and 9 of G09.  The
authors found that stellar population age and metallicity are nearly
independent of effective radius but strongly correlated with velocity
dispersion.  An analysis of the 6dFGS, which has a wider fiber
aperture than SDSS, found similar correlations
\citep{Magoulas:2012b,Springob:2012a}.  In all three slices of the FP,
galaxies with larger $\sigma$ had older ages and higher metallicities.
Stellar population age was also inversely correlated with residual
surface brightness, so that the youngest galaxies tend to fall above
the FP, in agreement with earlier observations
\citep{Forbes:1998a,Terlevich:2002a}.  A key conclusion of a
subsequent analysis \citep{Graves:2010a,Graves:2010b} was that these
trends arise because of structural differences in galaxies. These
papers speculated that galaxies below the FP may have had earlier
truncation times and formed most of their stars early, while galaxies
above the FP have more extended star formation histories.

\section {Results}
\subsection {Binning in the fundamental plane}
We make use of 23 of the $(50\ h^{-1} \rm Mpc)^3$ subvolumes of the
Bolshoi simulation in this analysis. We attempt to select a sample of
low-redshift passive spheroid-dominated galaxies that is as similar as
possible to the G09 sample. In order to do this, we select galaxies
with stellar mass $B/T>0.5$, specific star formation rates less than
0.1 \Mdot $\ \mathrm{yr}^{-1}/10^{11}$ \Mdot, and r-band absolute
magnitudes $M_{r} > -19.0$ (the G09 $50\%$ completeness threshold is
$M_{r} = -19.7$). Making these selection cuts, we obtain a sample of
\ngal \ model galaxies. 

We note that the population of simulated galaxies consists solely of
spheroid-dominated galaxies that have formed their spheroids via
mergers or disk instabilities; this may not exactly correspond to the
population of observed early-type galaxies.  Based on an analysis of
the SDSS, \citet{Cheng:2011a} have shown that a sample of passive, red
sequence galaxies selected based on {\color{black} an apparent}
$B/T>0.5$ (similar to the G09 sample) may {\color{black} actually}
contain a significant fraction of disk-dominated (passive S0 and Sa)
galaxies. These galaxies might be created via environmental processes
such as ram pressure stripping that are not included in our model. We
have made no attempt to exclude galaxies that the SAMs characterize as
having {\color{black} (``bright mode'')} AGN at redshift zero;
{\color{black} G09 excluded Seyferts from their sample to avoid
  contamination of the spectra with emission lines, but as long as the
  structural properties of AGN are no different from inactive
  galaxies, this should not induce a bias}.
%(these are rare).  
While the results presented here are for galaxies at redshift zero we
have also checked that including galaxies from the range $0 < z <
0.08$ does not significantly change the results.

\begin{figure*}
	\centering
			{\includegraphics[width=10cm]{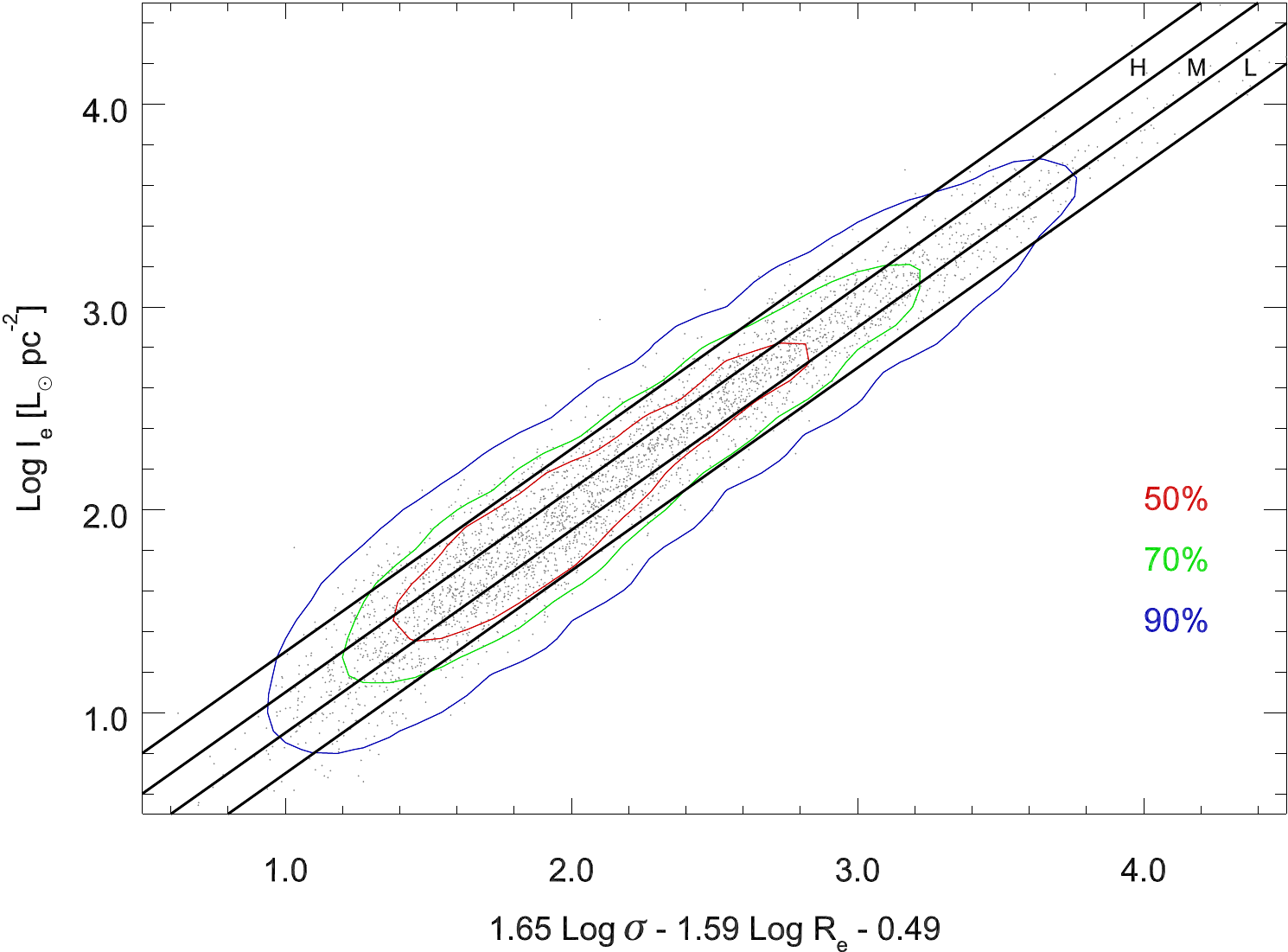}}
\caption{Distribution of simulated galaxies through the thickness of
  the fundamental plane.  Galaxies are fit to a linear relation
  (horizontal axis) relating surface brightness with velocity
  dispersion and radius.  The measured surface brightnesses are then
  plotted against the expected values.  The areas between the solid
  black lines represent the slices we term the `low-FP' (L),
  `midplane' (M), and `high-FP' (H), from bottom to top, according to
  the residual in surface brightness. Each slice has a thickness of
  0.2.  \SFitPercent \ of the galaxies fall within the middle three FP
  slices.  The red, green, and blue contours enclose 50\%, 70\%, and
  90\% of all galaxies, while the grey points represent individual
  galaxies.}
	\label{fig:FP_thickness}
\end{figure*}

We then separate our `early-type' sample into three regimes according
to their location perpendicularly above or below (i.e. `through') the
fundamental plane, using surface brightness as the independent
variable.  {\color{black} Surface brightness is calculated for our SAM
galaxies via the luminosity given by stellar population synthesis models.}
Since we intend to compare to the G09 results, only
galaxies that fall within the G09 range of radius, $\mathrm{(0.0 <
  \log(\emph{R}_{e}/ kpc) <0.7\ )}$ and velocity dispersion,
$\mathrm{(2.0\ < \log (\sigma/km \ s^{-1}) <2.4)}$ at redshift zero,
are included in the fitting routine.  We use a least-squares fit to
determine a relation between (log) $R_{\mathrm{e}}$, (log) $\sigma$,
and (log) $I_{\mathrm{e}}$, finding the fundamental plane prediction
\begin{fleqn}
	\begin{equation}
		\label{eqn:S12_FP_params} 
		\SFit.
	\end{equation} 
\end{fleqn}
For each galaxy, the predicted surface brightness is determined using
the above relation, and galaxies are separated by their residuals
$\Delta \log (\emph{I}_{e}/\mathrm{\Ldotpc})$. Residuals in the three ranges
[-0.3,-0.1], [-0.1,0.1], and [0.1,0.3] are respectively termed the low-, mid-, and
high-FP.  Galaxies `above' the FP have higher surface
brightnesses than one would predict using their effective radii and
velocity dispersions, while galaxies `below' the FP have lower surface
brightnesses.  Galaxies with residuals outside the range [-0.3,0.3]
are excluded.  If we plot the simulated galaxies according to their
location in FP-space, \SFitPercent \ of the galaxies fall within the
low-to-high FP slices (Figure \ref{fig:FP_thickness}).

After separating the galaxies by their location within the FP, we
place them in bins according to their radius and velocity dispersion
just as G09 did for SDSS galaxies.  We define this face-on projection
in $\mathrm{R_{e}}$ and $\sigma$ to be `across' the FP.  We then
calculate the median age and metallicity for all galaxies within each
bin, discarding bins with fewer than five galaxies.  These values are
used to form contours relating the stellar population parameters,
namely age and metallicity, with the fundamental plane parameters and
residuals.  We caution that the simulated quantities are
mass-weighted, while G09 calculates light-weighted ages,
metallicities, and effective radii.

\begin{figure*}
	\centering
		\subfigure{
			{\includegraphics[width=15cm]{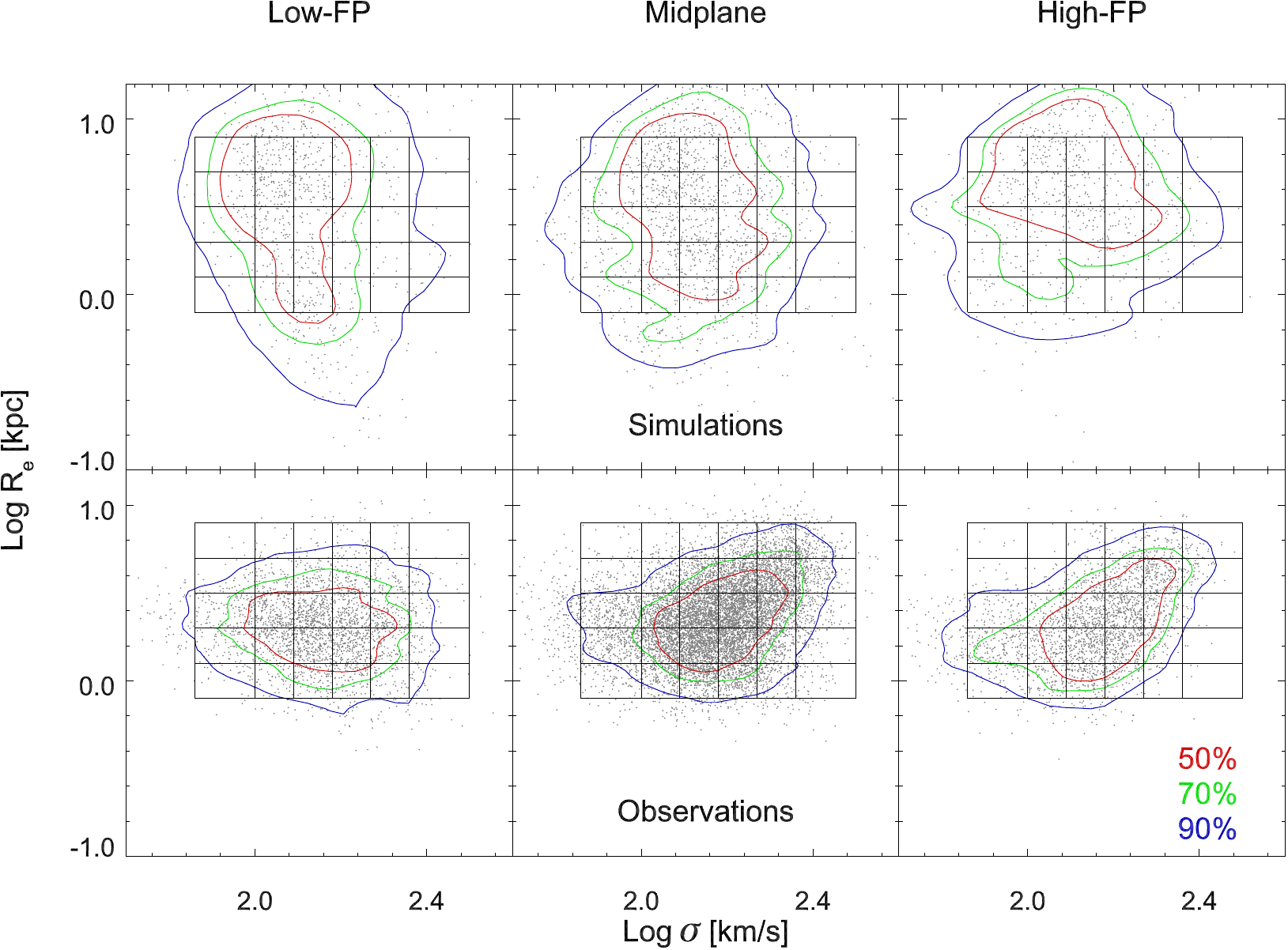}}
			}
\caption{Distribution of radius and velocity dispersion for galaxies
  within each slice of the FP for the P14 semi-analytic model (top)
  and G09 observations (bottom).  From left to right, the panels
  represent the `low-FP', `midplane', and 'high-FP' slices.  The grid
  lines show the bin definitions in the region of the G09
  observations; the median age and metallicities are calculated within
  each bin.  The SAM contains a population of galaxies with low radii
  and low surface brightnesses that fall below the completeness
  threshold of the G09 survey.  The red, green, and blue contours
  enclose 50\%, 70\%, and 90\% of galaxies meeting our selection
  criteria, while the grey points represent individual galaxies.}
			\label{fig:FP_bin}
	
\end{figure*}

Comparing the P14 SAM and G09 observed populations, we find that they
occupy slightly different regions of the $R_{\mathrm{e}}$-$\sigma$
parameter space (Figure \ref{fig:FP_bin}).  The SAM includes a
population of galaxies with low radii and low surface brightnesses
that is not seen in G09.  These galaxies tend to have low stellar
masses and absolute magnitudes, and fall below the G09
completeness threshold.

%%%Old Figure 3, Omitted
%\begin{figure*}
%	\centering
%			{\includegraphics[width=\lbox]{3.pdf}} 
%\caption{Relation between mass-weighted age (top) and metallicity
%  (bottom), effective radius, and velocity dispersion for
%  ``early-type'' galaxies in the P14 SAM.  The different panels
%  represent the three central slices of the FP, as shown in Figure
%  \protect\ref{fig:FP_bin}.  The grey dashed line indicates the region
%  analyzed in G09.  Stellar population age increases strongly with
%  velocity dispersion and has no clear trend with radius.  These
%  results are in qualitative agreement with observations.  Metallicity
%  increases with both radius and velocity dispersion (contours are
%  tilted); this is in contrast to the G09 observations, in which
%  metallicity increases solely with velocity dispersion.  At fixed
%  radius and velocity dispersion, galaxies that lie above the FP tend
%  to be younger and more metal-rich than those that lie below the FP.}
%					\label{fig:main_contours} 
%\end{figure*}

\begin{figure*}
	\centering
			{\includegraphics[width=\lbox]{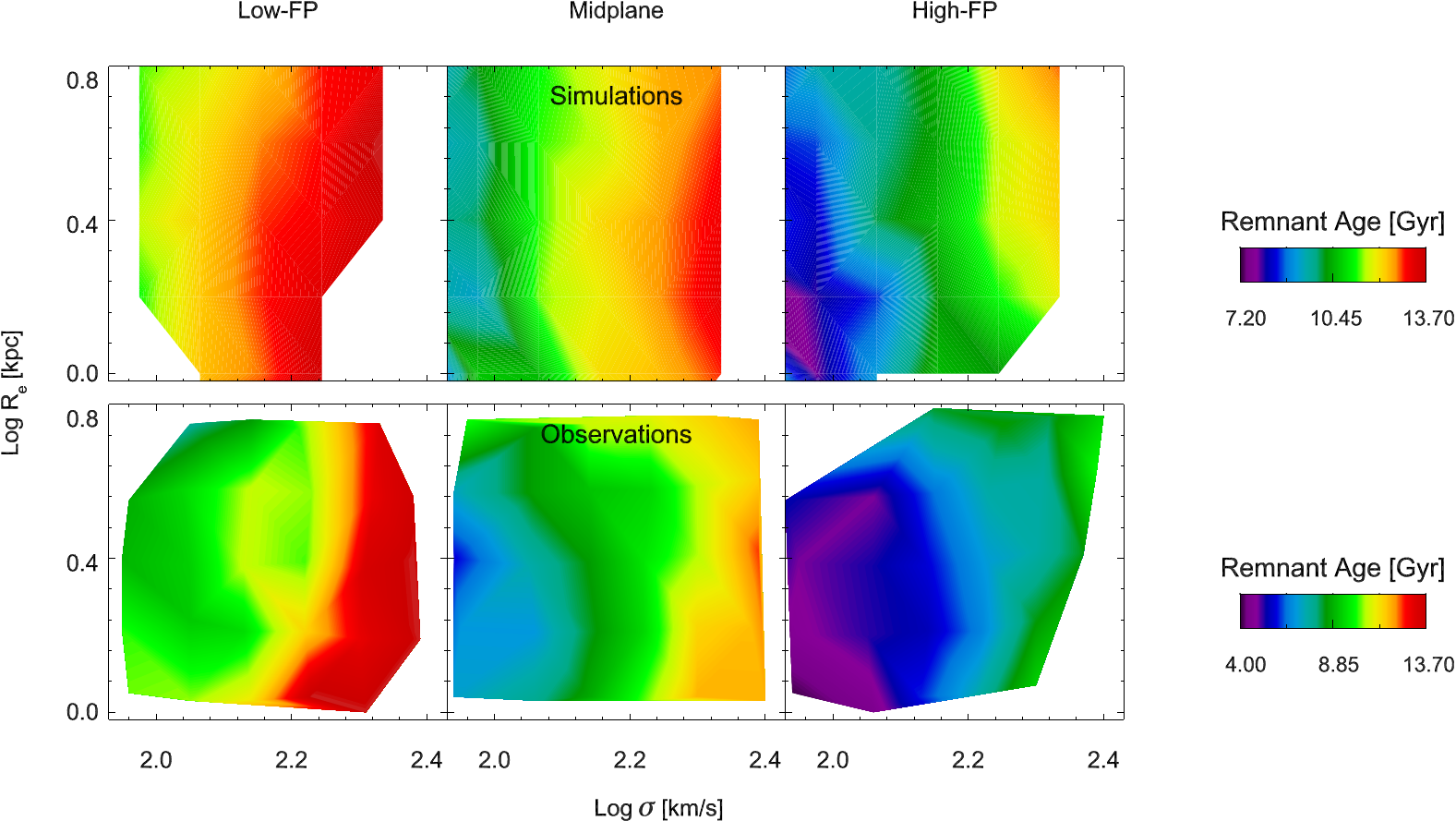}} 	
\caption{Relation between mass-weighted age, effective radius, and
  velocity dispersion for early-type galaxies in the P14 SAM (top) and
  G09 observations (bottom).  Here we plot only the region considered
  in G09.  The different panels represent the three central slices of
  the FP, as shown in Figure \protect\ref{fig:FP_bin}.  In the SAM and
  the observations, stellar population age increases with velocity
  dispersion, but the SAM galaxies display a narrower range in age.
  Galaxies that lie above the FP also tend to be younger than those
  that lie below the FP.}
					\label{fig:age_mid_contour}
	
\end{figure*}

\begin{figure*}
	\centering
		{\includegraphics[width=\lbox]{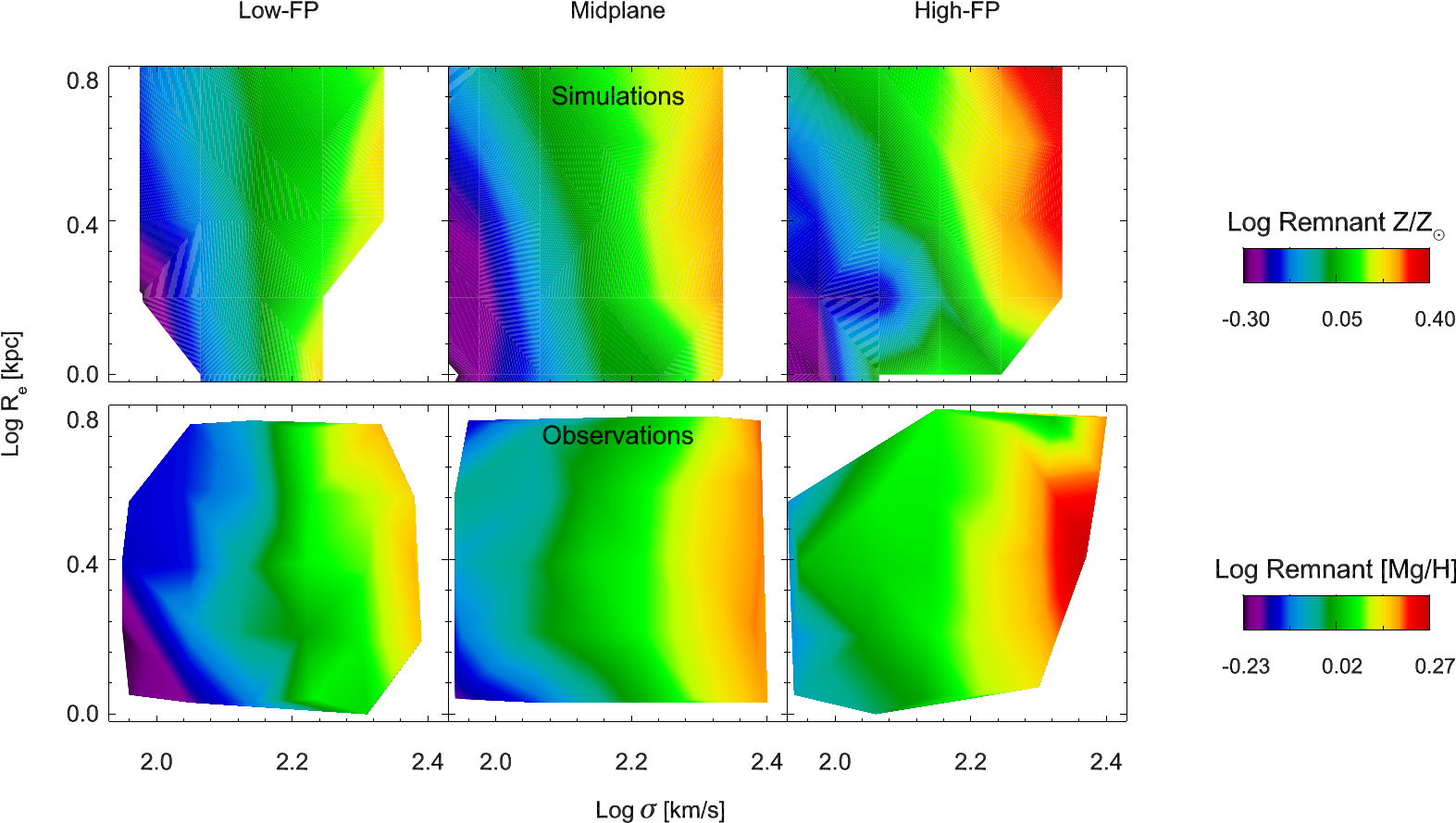}} 
  			
\caption{Relation between mass-weighted metallicity, effective radius,
  and velocity dispersion for early type galaxies in the P14 SAM (top)
  and G09 observations (bottom).  Here we plot only the region
  considered in G09.  The different panels represent the three central
  slices of the FP, as shown in Figure \protect\ref{fig:FP_bin}.
  While [Mg/H] depends strongly on velocity dispersion in G09, in the
  SAM metallicity depends on both velocity dispersion and effective
  radius.  The simulated galaxies tend to have slightly lower
  metallicities than observations on average.}
				\label{fig:me_mid_contour}
	
\end{figure*}

\begin{figure*}
	\centering
			{\includegraphics[width=\lfull]{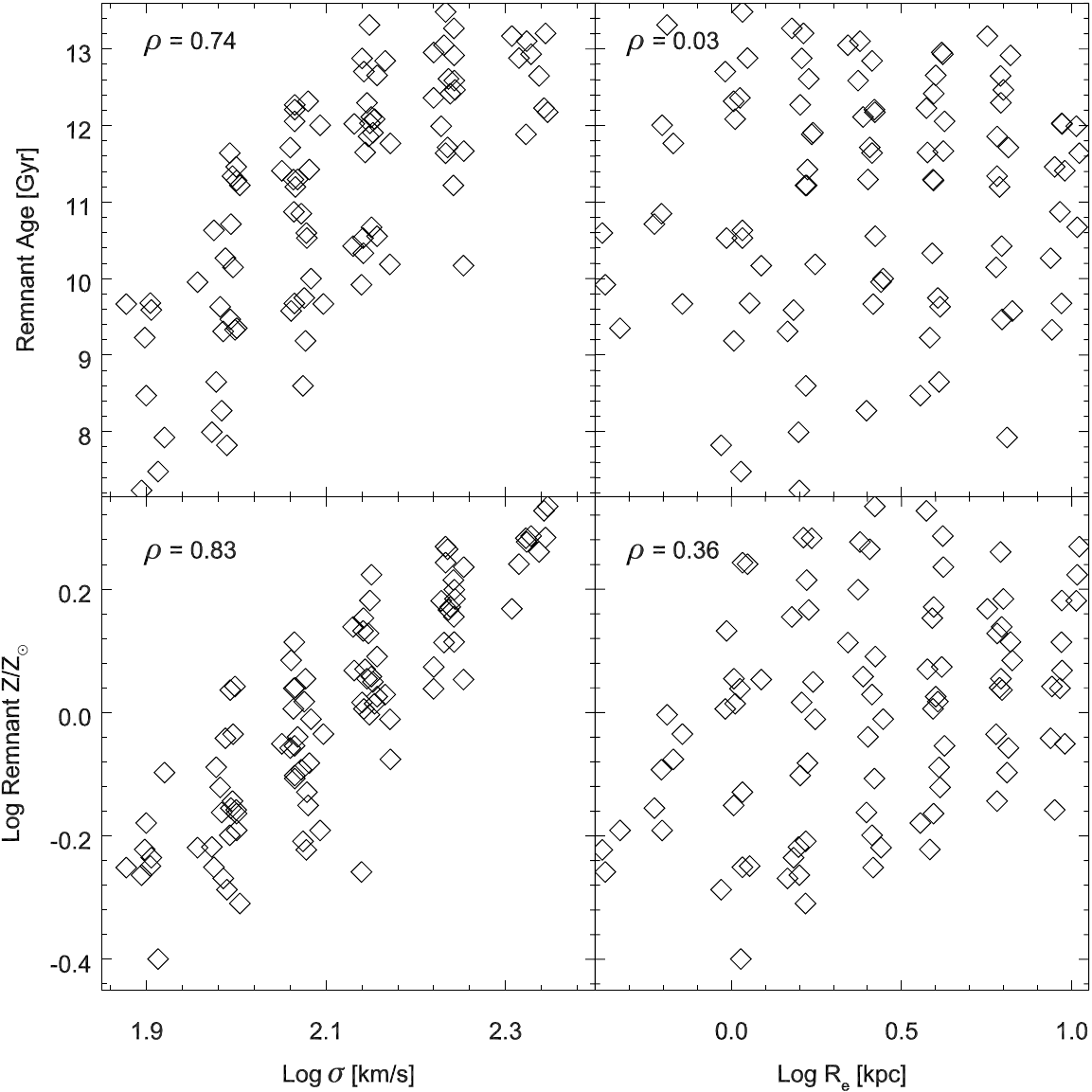}} 
\caption{Predicted relation between mass-weighted age (top) and
  metallicity (bottom), velocity dispersion (left) and effective
  radius (right), for each of the bins in the FP, for the P14 SAM.
  Each panel shows the Spearman rank coefficient $\rho$ indicating the
  strength of the two-dimensional relation.  Both age and metallicity
  are strong functions of velocity dispersion. Age is nearly
  independent of effective radius, while metallicity weakly increases
  with effective radius.}
					\label{fig:spearman} 
\end{figure*}

Our major findings are the contours seen {in the upper panels of Figures} 
%\ref{fig:main_contours}.  
\ref{fig:age_mid_contour} and \ref{fig:me_mid_contour}.
Stellar population age is positively
correlated with velocity dispersion and is only weakly dependent on
radius.  We note that the parameter space has a much larger range in
radius than velocity dispersion, so that while the contours appear
nearly vertical, the radial dependence is non-negligible.  If we
consider the the entire FP, the correlation between stellar age and
effective radius has a Spearman rank coefficient $\rho = $
\SpearmanAgeRe \, indicating nearly no correlation (Fig.~\ref{fig:spearman}).  The relationship
between age and velocity dispersion is much stronger, with a Spearman
rank coefficient $\rho = $ \SpearmanAgeSig.  Looking through the
thickness of the FP, galaxies that lie above the FP (those with the
largest residuals in $\mathrm{log}\ I_{\mathrm{e}}$) have younger
ages, as found by several observational studies
\cite[G09]{Forbes:1998a,Terlevich:2002a}.  Galaxies above the FP have
a mean age of $10.02 \pm 1.61 $\ Gyr, as compared to $12.12 \pm 1.12
$\ Gyr for galaxies below the FP.

%rss this is for the SAMs, right?
 	
If we compare the metallicity contours, the differences between the
SAM and observations are more pronounced.  In the simulated galaxies,
metallicity increases strongly with velocity dispersion ($\rho = $
\SpearmanZSig) and weakly with effective radius ($\rho = $
\SpearmanZRe), whereas the G09 galaxies show very little dependence of
metallicity on effective radius.  The predicted dependence on
effective radius is strongest for galaxies above the FP.  As in G09,
galaxies that lie above the FP do tend to have higher metallicities
($[Z]= 0.11 \pm 0.52$ vs. $[Z] = 0.01 \pm 0.17$ for high-FP and low-FP
galaxies, respectively).

\subsection{Analysis of the age and metallicity trends}

The strong correlation between metallicity and velocity dispersion in
the SAM is not unexpected; it arises from the dependence {\color{black} 
on galaxy circular
velocity} of the gas and metal ejection rate due to stellar feedback 
(see S08 and P14). In fact the
scatter in the mass-metallicity relationship predicted by the SAM for
the whole galaxy population is much smaller than that measured by
\citet{Gallazzi:2005a} {\color{black} for low-mass galaxies}, 
though a smaller scatter is obtained in other
observational studies using more direct metallicity indicators
\citep{Woo:2008a,Kirby:2010a}. 
%Furthermore, we have presented
%mass-weighted metallicities for the simulated galaxies, compared with
%effectively light-weighted metallicities for the observations.

\begin{figure*}
	\centering
   		\subfigure{
			{\includegraphics[width=12cm]{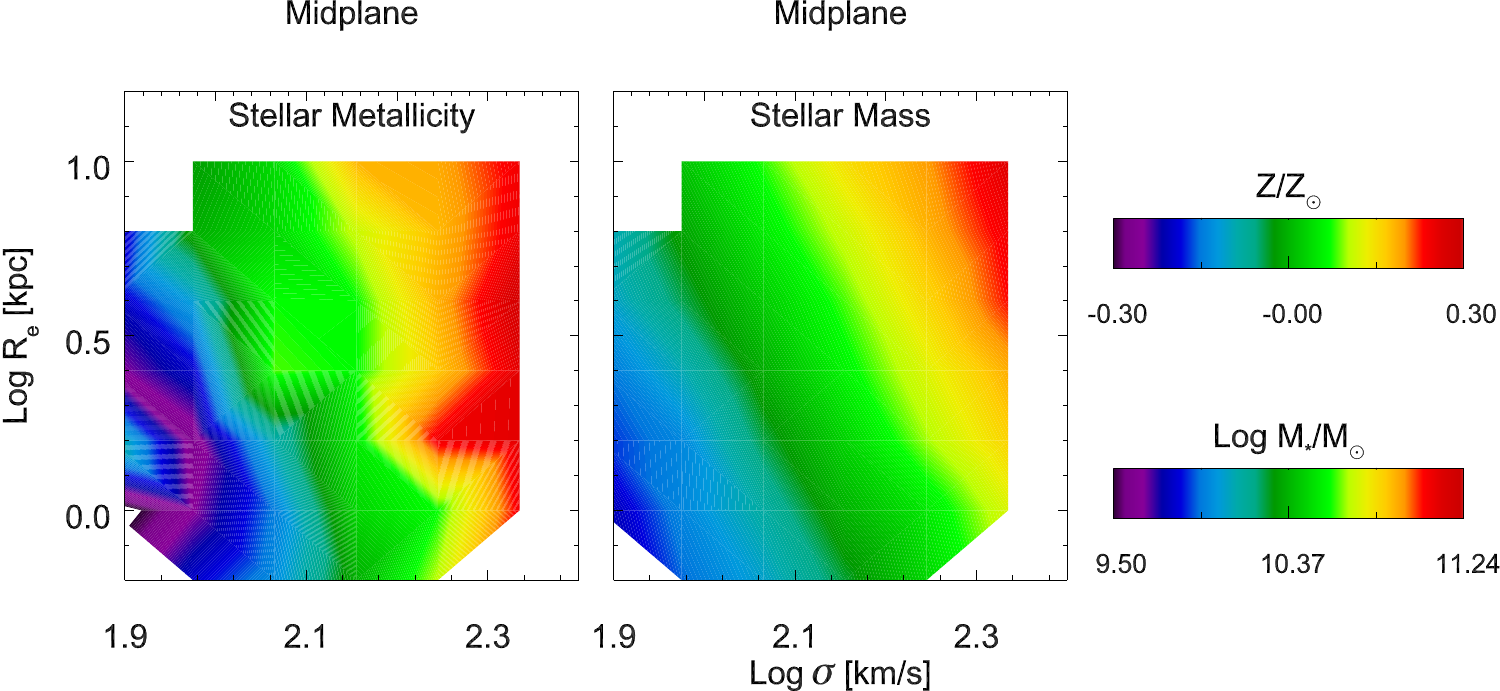}} 
			}
\caption{Relation between mass-weighted metallicity (left) and stellar
  mass (right), effective radius, and velocity dispersion for early
  type galaxies in the P14 SAM.  Both panels represent the middle
  slice of the FP, as shown in Figure \protect\ref{fig:FP_bin}.
  Colors are individually normalized.  Since the SAM predicts a very
  strong correlation between stellar mass and metallicity, the two
  trends are nearly identical.  }
\label{fig:mass_metallicity_contour} 
\end{figure*}

If we consider that stellar mass is closely related to the dynamical
mass, which is in turn proportional to $\sigma^{2}r$ then the
implications of the upper panels of Figures %\ref{fig:main_contours} 
\ref{fig:age_mid_contour} and \ref{fig:me_mid_contour}
become clearer: the
tight mass-metallicity relationship is reflected in a dependence of
metallicity on both effective radius and velocity dispersion in the
projected fundamental plane.  In fact, when stellar mass and
metallicity are plotted alongside each other in the middle slice of
the fundamental plane (Figure \ref{fig:mass_metallicity_contour}), 
we find that the trends are %nearly identical
{\color{black} similar, although metallicity is relatively independent of $R_e$
at high $\sigma$, whereas higher-$R_e$ galaxies have higher
stellar masses at high $\sigma$}.  As we will show,
galaxies above the fundamental plane have the lowest concentrations of
dark matter within their effective radii, making the dependence on
stellar mass even more pronounced.

We next discuss the simulated age-FP trend, which shows {\color{black}
a better} agreement with observations {\color{black} than might have
been expected}.  We note that, in the model, the velocity
dispersion is calculated from both the effective radius and the total
mass within that radius; thus velocity dispersion and effective radius
are intrinsically linked.  However, the ages of the simulated galaxies
have a clear dependence on the former and %almost no 
{\color{black} less} dependence on the
latter.  As we will discuss in section 5.1, minor mergers are mainly
responsible for blurring any correlation between age and effective
radius while preserving the relation between age and velocity
dispersion.

\subsection{Comparison to observations}
To %better
compare with G09, we have replotted the age and metallicity
contours over the range $\mathrm{(0.0 < \log(\emph{R}_{e}/ kpc)
  <0.7\ )}$, $\mathrm{(1.9\ < \log (\sigma/km \ s^{-1}) <2.4)}$, and
$-0.3 <\Delta \log (\emph{I}_{e}/\mathrm{\Ldotpc}) < 0.3$ considered
by \cite{Graves:2009b} alongside the G09 data (Figures
\ref{fig:age_mid_contour} and \ref{fig:me_mid_contour}).  We caution
that the G09 ages were later found to be systematically high by $\sim
0.12$ dex, owing to weak emission in the H$\beta$ absorption line
\citep{Graves:2010a}; however, this should not affect the overall
trends.  We also note that we calculate mass-weighted ages and
metallicities while G09 calculated single stellar population (SSP)
ages and metallicites using the Lick indices.  These SSP quantities
have been shown to more closely correlate with the epoch of most
recent star formation, and result in ages that are systematically
younger than mass-weighted and light-weighted ages
\citep{Trager:2009a}.
  
Examining the trends within FP slices, the age-FP correlations are in
rough agreement with G09.  The major difference between our results
and those of G09 is that we find metallicity to be dependent on radius
and velocity dispersion, especially above the FP, while G09 found
metallicity to be dependent on velocity dispersion alone. The SAM does
a better job of reproducing the observed trends through, as opposed to
across, the FP.  Galaxies that fall above the FP tend to be younger
and more metal-enhanced than average, while those that fall below the
FP are older and more metal-poor, in agreement with G09.

\section{Discussion}

%In the previous section, we determined that less evolution  galaxy's velocity dispersion at high redshifts and at redshift zero.  This finding is supported by several recent observations \citep{Cenarro:2009a, B is An important implication of 
Having established the major trends of age and metallicity through the
thickness of the fundamental plane, we can now attempt to characterize
the significance of these trends.  A key question is how much of the
variation through the fundamental plane arises from structural
differences in galaxies as compared to the passive fading of
elliptical galaxies.  If galaxies do `settle' onto the FP over time,
we might expect galaxies above the FP to have younger ages and higher
metallicities, in agreement with both simulations and observations.
However, this process would not explain why age and metallicity appear
to be more strongly correlated with velocity dispersion than effective
radius.  In addition, this would not explain the significant overlap
in age and metallicity ranges in the three FP slices.

\subsection{Analysis of trends across the FP}
In order to understand all of these trends simultaneously, it is
necessary to study the implications of our prescription for effective
radius and velocity dispersion.  In our model, a galaxy's effective
radius and velocity dispersion are tightly correlated, regardless of
whether it forms a spheroid through a merger or through a disk
instability event.  As a population however, galaxies experience large
changes in effective radius with redshift, but only moderate changes
in velocity dispersion.  By quantifying the scatter introduced by
these evolutionary processes we can attempt to explain how
low-redshift galaxies with similar ages and metallicities have similar
velocity dispersions but a range of effective radii.

Recent works \citep{Naab:2009a,Hopkins:2010b,Oser:2012a} have
suggested that gas-poor minor mergers may produce at least some of the
observed evolution in the size-mass relation for early-type galaxies,
forming an evolutionary link between the compact galaxies seen at high
redshifts and the more diffuse galaxies seen in the local universe.
While there is some question as to whether the merger rate is
sufficient to explain all of the size evolution
\citep{Trujillo:2011a,Newman:2012b,Nipoti:2012a,Quilis:2012a}, we can
predict the effect that these events would have on the population.

It is important to note that in our model, following the behavior in
the numerical simulations, any merger where one or both of the
progenitors is spheroid-dominated is treated as a dissipationless event,
as we explained in Section 2.2.  Thus, mergers between a massive
compact elliptical and a smaller galaxy can be expected to
significantly increase the size of the remnant galaxy.  Using
conservation of energy and the virial theorem, \cite{Naab:2009a} show
that for a series of minor mergers that increase the mass of the
galaxy from $M_{\mathrm{i}}$ to $M_{\mathrm{f}}$, the radius increases
as $\left(M_{\mathrm{f}}/M_{\mathrm{i}}\right)^{2}$ while the velocity
dispersion decreases as
$\left(M_{\mathrm{f}}/M_{\mathrm{i}}\right)^{-1/2}$.  These scaling
relations necessarily introduce a large amount of scatter in effective
radius: if, for example, two identical galaxies increase their masses
by a factor of 1.9 and 2.0 respectively, their resulting radii will
differ by 9.8\% while their velocity dispersions will only differ by
2.6\%.

This large amount of variation in effective radius means that any
original correlations between effective radius and age or metallicity
will be weakened by a series of minor mergers.  It is interesting to
note that this model predicts that there may be a stronger dependence
of stellar population parameters on effective radius at higher
redshifts, where the effects of minor mergers are less prevalent.

A second major implication of these scaling relations is that the
velocity dispersion of a galaxy should remain relatively unchanged
from its formation to the present day; if anything, it should decrease
slightly.  This prediction is in agreement with both cosmological
simulations \citep{Oser:2012a} and observational evidence that the
velocity dispersion function evolves to higher values at higher
redshifts but at a rate much slower than the evolution in the
size-mass relation \citep{Cenarro:2009a,Bezanson:2011a}.

\begin{figure*}
	\centering
			{\includegraphics[width=\lbox]{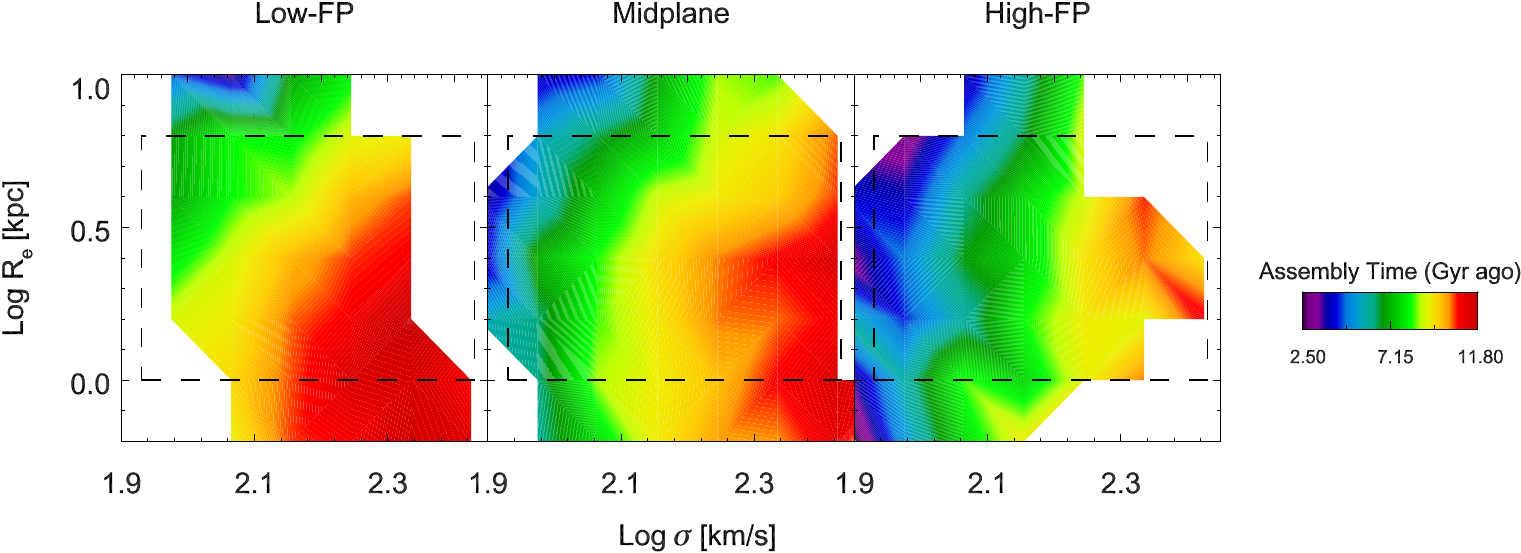}}
\caption{Relation between the time since the galaxy was assembled
  (became spheroid dominated), effective radius, and velocity
  dispersion for early type galaxies.  The different panels represent
  the three central slices of the FP, as shown in Figure
  \protect\ref{fig:FP_bin}.  Early type galaxies that were ssembled
  earlier have higher velocity dispersions and tend to fall below the
  FP.}
					\label{fig:tassemble_contour}
	
\end{figure*}

Since our SAM, based on the Bolshoi simulation merger trees, contains
the detailed merger history of every simulated galaxy, we are able to
test this prediction directly.  If we define the `assembly time' as
the time a galaxy most recently became spheroid-dominated ($B/T>0.5$),
we can examine its variation across and through the FP using the same
method as described earlier for age and metallicity.  The results can
be seen in Figure \ref{fig:tassemble_contour}.  As expected, galaxies
with higher velocity dispersions tend to have assembled earlier.
Thus, a galaxy's velocity dispersion may be a key indicator relating
its current structure to the epoch of its formation.

Figure \ref{fig:tassemble_contour} also shows that galaxies below the
FP tend to have earlier assembly times.  There is a significant amount
of overlap in this correlation however; in particular the galaxies
with the highest velocity dispersions have similar high formation
times for all three slices of the FP.  This indicates that a further
process must be invoked to explain the trends through the thickness of
the FP.

\subsection{Analysis of trends through the FP}
Stellar population trends through the thickness of the FP can arise in a
number of different ways; any process that increases the dynamical
mass-to-light ratio of a galaxy will move it further below the virial
plane.  \cite{Graves:2010a} provide a decomposition of the deviation
from the virial theorem, separating it into four components:
\begin{enumerate}
\item The ratio of the estimated dynamical mass to the true mass within one $\mathrm{R_e}$.
\item The ratio of the true mass within one $\mathrm{R_e}$ to the projected stellar mass.
%\item The ratio of the projected stellar mass with an assumed initial mass function (IMF) to the true stellar mass and its corresponding IMF.
\item The ratio of the projected stellar mass within one $\mathrm{R_e}$ to 
the stellar mass computed with an assumed IMF.
\item The stellar mass-to-light ratio for the assumed IMF.
\end{enumerate}  
For the simulated galaxies and their corresponding FP, the first and
third terms are identically one, as we have no uncertainty in the
dynamical mass estimate and we model and `observe' galaxies using the
same IMF.  Since we know the stellar mass-to-light ratio for the
galaxies and can calculate the central dark matter fraction (DMF, see
below), we may calculate the second and fourth terms directly. We note
that if our assumed Chabrier IMF is incorrect, or if the IMF is
non-universal \citep{Conroy:2012a,Dutton:2012b,Spiniello:2012a} the
fourth term would change; we discuss the implications of a non-universal
IMF later in this section.

We have used the same process as described above to project the
dynamical-to-stellar mass and stellar mass-to-light ratios of
early-type galaxies across (i.e., along the face-on projection) and
through the FP (Figure \ref{fig:mass_contour}).  The results 
{\color{black} regarding variations through the FP are in
agreement with the conclusions of \cite{Graves:2010a}: galaxies that
fall below (above) the FP have higher (lower) dynamical-to-stellar
mass ratios and slightly higher (lower) stellar mass-to-light ratios,
{\color{black} at fixed velocity dispersion and $\mathrm{R_{e}}$.}
Stated another way, galaxies below the FP have lower stellar masses
and central surface densities at fixed $\mathrm{R_{e}}$.  The
variations in stellar mass-to-light ratio are due to differences in
the stellar populations: since galaxies above the FP are younger than
galaxies below the FP, they have more young stars and hence %higher 
{\color{black} lower} stellar mass-to-light ratios.  The variations in the central dark
matter fraction reflect structural differences in the density profiles
of galaxies and dark matter halos.  

Comparing the trends through the thickness of the FP, at fixed
$\mathrm R_{e}$ and $\sigma$ the dynamical-to-stellar mass ratio has a
much larger degree of variation than the stellar mass-to-light ratio.
If we limit our analysis to bins of $\mathrm{R_{e}}$ and $\sigma$ that
have at least 5 galaxies in each slice of the FP, we find that the
average variance in the dark matter fraction contributes 94\% of the
thickness of the FP, while the stellar mass-to-light ratio only
contributes 6\%.  This is in general agreement with
\cite{Graves:2010a}, who found that the dark matter fraction and IMF
variation have a combined contribution in the range of 47\% - 98\%,
and that measured variations in the stellar mass-to-light ratio were
insufficient to explain all of the thickness of the FP.  This is
another indication that underlying structural differences, as opposed
to passive fading, are the main contributors to the thickness of the
FP.

We note that our model assumes that all stars formed under a Chabrier
IMF.  There is mounting evidence, however, that the IMF may be
non-universal.  Early-type galaxies and spheroids with high stellar
masses or velocity dispersions in the local universe may follow a
`bottom-heavy' IMF, with more low-mass stars
\citep{Conroy:2012a,Dutton:2012b,Spiniello:2012a}.  While there is
some disagreement as to the slope of this bottom-heavy IMF, an IMF
that varies with velocity dispersion would contribute to the thickness
of the simulated FP; galaxies below the FP with high velocity
dispersions would have higher stellar mass-to-light ratios, moving
them even further below the FP.  Thus while our results {\color{black}
through the FP} are consistent
with those of \cite{Graves:2010a}, we have not accounted for any
contributions from a varying IMF in this work.
{\color{black} Regarding the variation in $M_*/L$ and $M_{dyn}/M_*$
across the FP, we note that although \citet{Graves:2010a} find that 
$M_*/L$ varies with $\sigma$ and is nearly independent of $R_e$
in agreement with Figure \ref{fig:mass_contour} (bottom), they find that
$M_{dyn}/M_*$ increases with both $R_e$ and $\sigma$ while in 
Figure \ref{fig:mass_contour} (top) we find that $M_{dyn}/M_*$ mainly
increases with increasing $R_e$, with $M_{dyn}/M_*$ slightly 
decreasing with increasing $\sigma$.}

\begin{figure*}
	\centering
			{\includegraphics[width=\lbox]{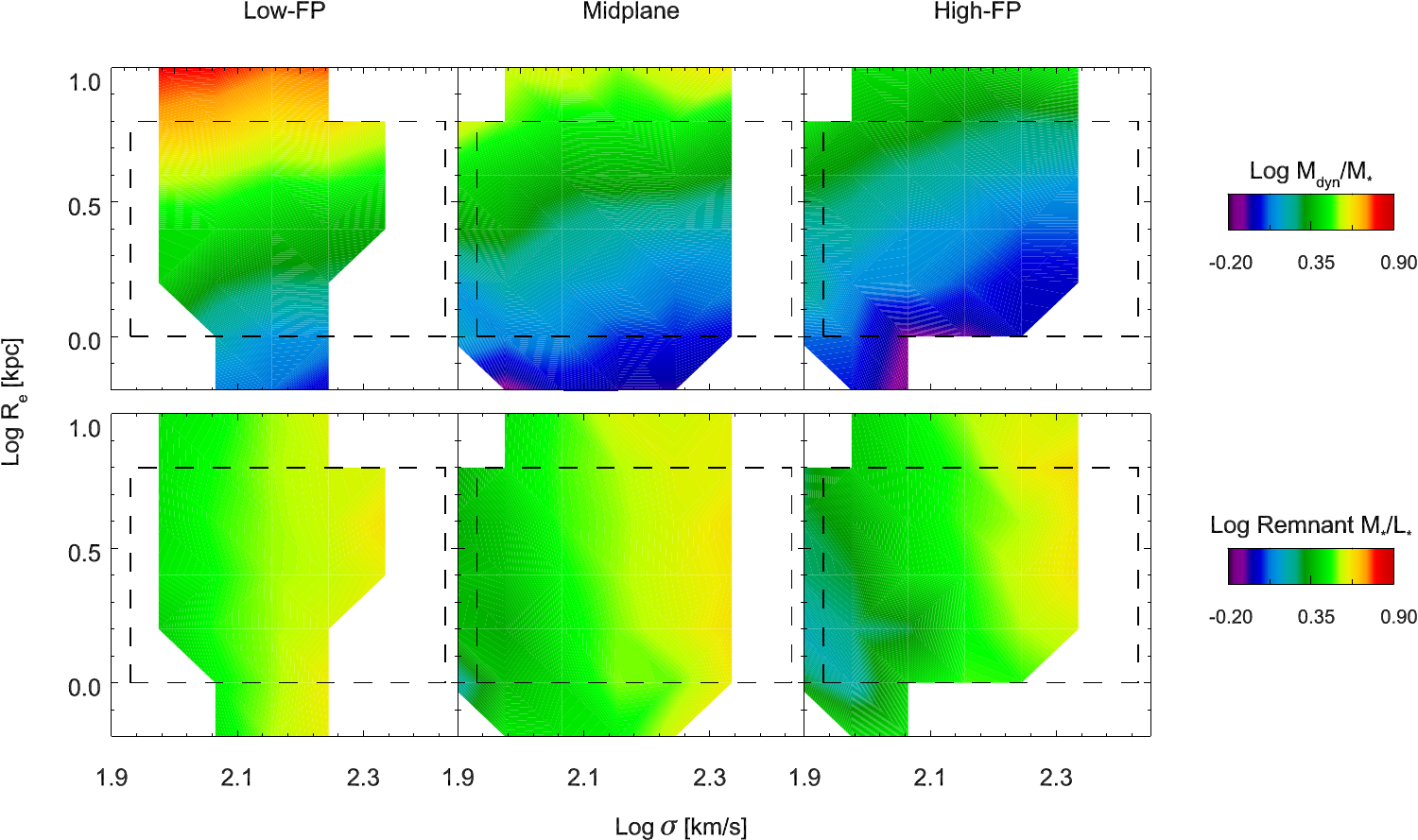}}
\caption{Relation between dynamical-to-stellar mass ratio (top),
  stellar mass-to-light ratio (bottom), effective radius, and velocity
  dispersion for early type galaxies in the P14 SAM.  The different
  panels represent the three central slices of the FP, as shown in
  Figure \protect\ref{fig:FP_bin}.  The grey dashed line indicates the
  region analyzed in G09. Galaxies that fall below the FP have higher
  dynamical-to-stellar masses and mass-to-light ratios.  The variation
  in dynamical-to-stellar mass through the FP is much larger than the
  variation in the mass-to-light ratio (note that both relations use
  the same color scalings).}
					\label{fig:mass_contour}
	
\end{figure*}

To summarize our results {\color{black} through the FP}
so far, we have found that galaxies below the
FP tend to be old and metal-poor.  They became spheroid-dominated at
early times, and have high dynamical-to-stellar mass ratios at a given
$\mathrm{R_{e}}$ and $\sigma$.  In contrast, galaxies above the FP
tend to be young and metal rich, with later formation times and
relatively high stellar masses at fixed $\mathrm{R_{e}}$ and $\sigma$.
In an analysis of the trends found in G09 and \cite{Graves:2010a},
\cite{Graves:2010b} found these results could be explained by a
scenario in which galaxies below the FP have their star formation
truncated at early times, while those above the FP have more extended
star formation histories.  Since the SAM contains information about
the star formation histories for every galaxy, we can test this
scenario.

The P14 SAM keeps track of the star formation history of each galaxy,
divided into 196 log-spaced bins in age.  For this analysis, we have
defined the `star formation duration' as the duration over which each
galaxy formed the middle 68\% of its stars.  Thus galaxies with a
wider distribution of stellar ages will have longer star formation
durations.  We have defined `formation time' as the time by which half
of the stars in the galaxy have formed; this quantity is significant
because it incorporates information about the stars that formed in
situ in the galaxy as well as those that were accreted.  Taken
together, the `assembly time' (Figure \ref{fig:tassemble_contour}) and
the `formation time' provide a link between the structure of a galaxy
and the properties of its stellar population.  These definitions are
summarized in Table 1.

\begin{table*}
\begin{center}
\caption{Parameters used to determine the mass assembly histories of
  early-type galaxies.  The low-FP, mid plane, and high-FP values are
  the mean of each parameter for early-type galaxies in each plane.}
% title name of the table
\centering
%\begin{tabular}{l c c c c}
\begin{tabular}{p{4cm} p{6cm} p{1.7cm} p{1.7cm} p{1.7cm}}
\hline\hline
Parameter & Description & Low-FP & Midplane  & High-FP
\\ [0.5ex]
\hline

%\hline
\hline Assembly time (Gyr ago) &Time when the galaxy most recently became spheroid-dominated ($B/T>0.5$) &
9.10 $\pm$ 2.29  &
  8.35$\pm$ 2.62  &
   6.65$\pm$ 3.30  \\
Formation time (Gyr ago) & Time by which 1/2 of the stars in a galaxy formed&
 10.88$\pm$ 0.77  &
  10.34$\pm$ 0.97  &
   9.28$\pm$ 1.37   \\
Star formation duration (Gyr) & Time to form the middle 68\% of a galaxy's stars &
 2.32$\pm$ 0.93  &
  2.82$\pm$ 1.04  &
   3.94$\pm$ 1.29  \\
\hline % inserts single-line
\hline
\end{tabular}
\label{tab:model}
\end{center}
\end{table*}
We have plotted the correlations between star formation duration and
formation time, effective radius, and velocity dispersion for all
three FP slices in Figure \ref{fig:tstars_contour}.  Comparing the
relations, it is evident that galaxies below the FP have early
formation times and short star formation timescales; in fact roughly
75\% of these galaxies have star formation timescales less than 2 Gyr.
Galaxies above the FP formed their stars slightly later, but more
significantly, they have much longer star formation timescales;  70\%
of these galaxies have star formation durations \emph{greater} than 2
Gyr.

The long duration of star formation allows for more conversion of gas
to stars, decreasing the dynamical-to-stellar mass ratio.  As we
established earlier that variations in the mass-to-light ratio cannot
account for most of the thickness of the FP, these galaxies will not
`fall' onto the FP over time.  Their assembly histories have produced
galaxies with high baryon fractions and high stellar central surface
densities, and their relatively recent star formation has produced
stellar populations with young ages and high metallicities.  This
result is a key prediction of our model: \emph{the thickness of the FP
  appears to be due to structural differences in the galaxies
  resulting from their differing formation histories}.  Previous works
\citep[e.g.,][]{Wechsler:2002b} have linked the $z=0$ concentrations of dark
matter halos with the duration and epoch of mass assembly; here we
find similar results for early-type galaxies.

\begin{figure*}
	\centering
			{\includegraphics[width=\lbox]{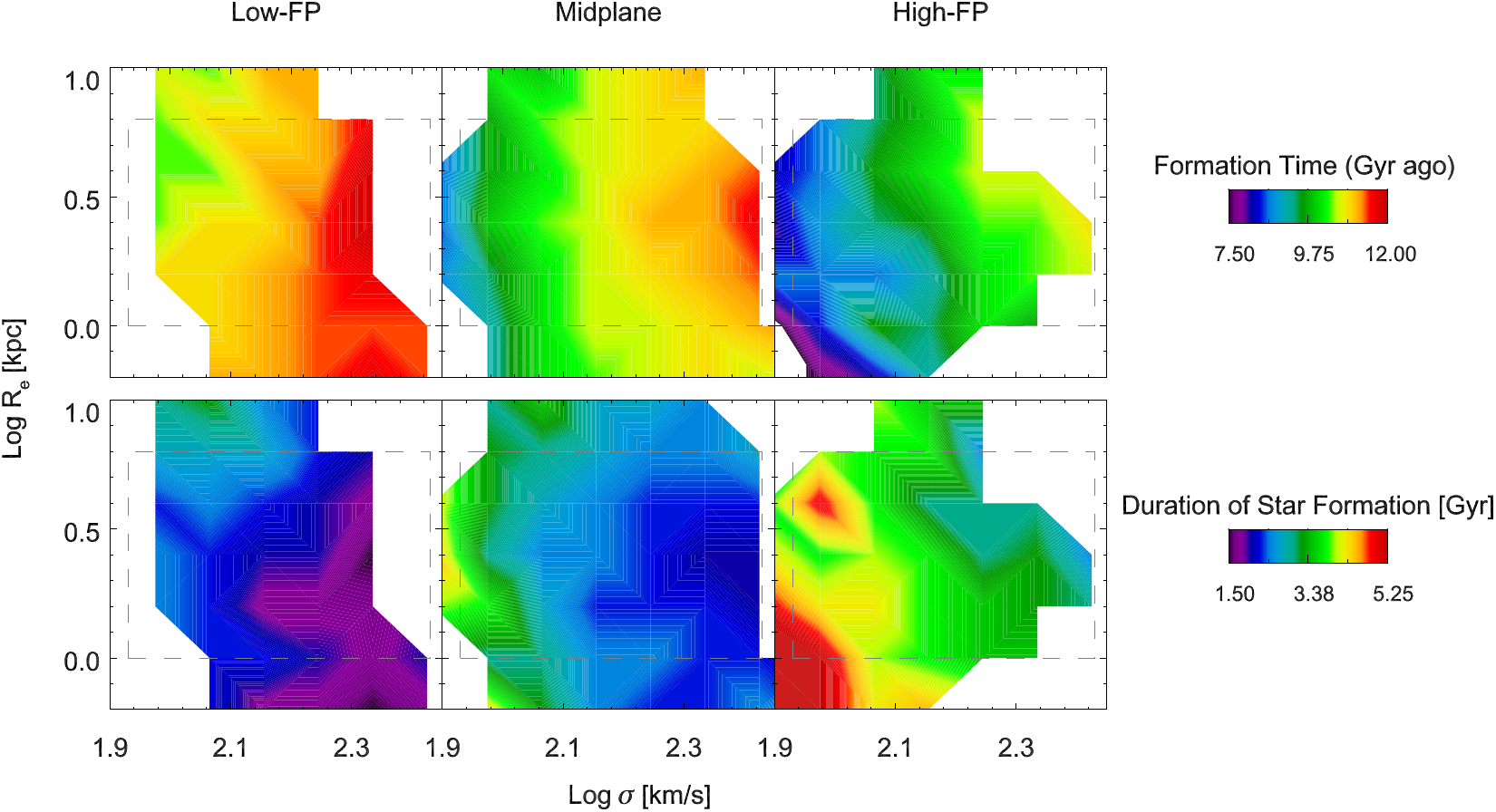}}
\caption{Relation between the formation time (top), duration of star
  formation (bottom), effective radius, and velocity dispersion for
  early type galaxies in the P14 SAM.  The different panels represent
  the three central slices of the FP, as shown in Figure
  \protect\ref{fig:FP_bin}.  The grey dashed line indicates the region
  analyzed in G09. Galaxies below the FP tend to have formed their
  stars early and have short star formation timescales.  Galaxies
  above the FP have extended star formation histories and formed their
  stars more recently.}
\label{fig:tstars_contour}
	
\end{figure*}

Finally, we can combine these relations to account for both the
structural and the stellar population differences through the FP.
Galaxies below the FP became spheroid-dominated early, in a regime in
which stellar velocity dispersions were higher at fixed stellar mass.
This may also account for the compact tail of galaxies seen in the
low-FP pane Figure \ref{fig:FP_bin}.  These galaxies formed their
stars and quenched early, leaving them with old ages, low
metallicities, and structural properties that are perhaps
representative of compact ellipticals at higher redshifts.

In contrast, galaxies above the FP became spheroid-dominated and formed
their stars slightly later.  More importantly, they have extended star
formation histories, producing galaxies with younger ages and higher
metallicities.  While they do have slightly lower stellar
mass-to-light ratios, most of the variation in residual surface
brightness stems from their high central stellar surface densities and
low dark matter fractions.

\section{Conclusions}

We have used the Santa Cruz SAM \citep[Porter et
  al. 2014]{Somerville:2008a,Somerville:2012a} along with an analytic
model for computing the size and velocity dispersion of stellar
spheroids from Covington et al. (2008, 2011) and \cite{Guo:2011a} to
predict the distribution of stellar ages and metallicities for
early-type galaxies across and through the fundamental plane. We allow
the model to run self-consistently to redshift zero, at which point we
select quiescent spheroid-dominated galaxies.  We then separate them
according to their residual surface brightness in the fundamental
plane and calculate the mass-weighted ages and metallicities as a
function of effective radius and velocity dispersion.

In agreement with G09 and an analysis of the 6dFGRS
\citep{Magoulas:2012b,Springob:2012a}, we find that stellar ages
increase as a strong function of velocity dispersion and are nearly
independent of radius.  We predict that the strong correlation with
velocity dispersion stems from the fact that the velocity dispersion
of the galaxy changes little from its formation to the present day,
even in the face of minor mergers.  Meanwhile minor mergers and
ongoing disk instabilities introduce large amounts of variation in the
radius over time, washing out correlations with effective radius.

We show that galaxies with higher residual surface brightness
(``above'' the FP) tend to be younger and more metal-rich.  Examining
their structural properties, we find them to have lower stellar
mass-to-light ratios and lower dynamical-to-stellar mass ratios.
These galaxies became spheroid-dominated relatively recently and formed
their stars later than galaxies below the FP.  Furthermore these
galaxies have extended star formation histories, allowing for a more
complete conversion of gas to stars and for the production of young,
metal-rich stars.  These results are in close agreement with the
observational analysis of \cite{Graves:2010b}, which also showed a
correlation between the chemical abundance of $\alpha$ elements, the
duration of star formation, and velocity dispersion.  A different
version of the Somerville et al. SAM contains a detailed Galactic
Chemical Evolution model, including non-instantaneous recycling,
enrichment by both core collapse and Type Ia supernovae, and tracking
of multiple chemical elements, as described in \cite{Arrigoni:2010a}.
This version of the SAM was shown to reproduce the observed scaling
between $[\alpha/\mathrm{Fe}]$ and velocity dispersion, suggesting
that our model would also reproduce the observed correlations between
$[\alpha/\mathrm{Fe}]$, the duration of star formation, and residual
surface brightness, as found in \cite{Graves:2010b}. We plan to pursue
this further in future work.  Note that \citet{Yates:14} also found a correlation
between stellar mass and $[\alpha/{\rm Fe}]$ for early-type galaxies
in a SAM, without any modification to the IMF.  Differences in star formation
timescale were also found to be a key cause in that work.

We find that variations in the stellar mass-to-light ratio and the
dark matter fraction within one effective radius both contribute to
the thickness of the FP, with the dark matter fraction having a much
larger effect.  Thus we predict that the thickness of the FP is
largely due to structural differences between galaxies, rather than
stellar population differences. Galaxies above the FP have higher
ratios of stellar-to-dark matter within one effective radius; put
another way, at fixed halo mass, galaxies above the FP have had more
efficient star formation.  This result is also in agreement with the
conclusions of G09, although we have not allowed for any contribution
from a non-universal IMF.  
The reasonably good agreement of our SAM
predictions with SDSS observations provides motivation to pursue more
detailed modeling.

\section{Acknowledgments}

We thank Matthew Colless, Brad Holden, Patrik Jonsson, Mark Krumholz,
Thorsten Naab, Ludwig Oser, Stefano Profumo, and Connie Rockosi for
useful discussions, {\color{black} and we thank the anonymous referee for many 
helpful questions and suggestions.}   
LAP thanks the Space Telescope Science Institute
for support and hospitality.  LAP and JRP were supported by NSF
AST-1010033 and the CANDELS grant HST-GO-12060.12-A.
\bibliographystyle{monthly}
%\bibliography{/Users/Lauren/Research/Writing/Support/Bibliographies/Combined.bib}
%\bibliography{/Users/rachel/work/research/mypapers/porter/Bibliographies/Combined.bib}
\bibliography{Combined.bib}

\begin{thebibliography}{100}
\expandafter\ifx\csname natexlab\endcsname\relax\def\natexlab#1{#1}\fi

\bibitem[{Adelman-McCarthy {et~al}\mbox{.}(2008)Adelman-McCarthy, Ag{\"u}eros,
  Allam, Zehavi, \& Zucker}]{Adelman-McCarthy:2008a}
Adelman-McCarthy J.~K., Ag{\"u}eros M.~A., Allam S.~S., Zehavi I., Zucker
  D.~B., 2008, ApJS, 175, 297

\bibitem[{Arrigoni {et~al}\mbox{.}(2010)Arrigoni, Trager, Somerville, \&
  Gibson}]{Arrigoni:2010a}
Arrigoni M., Trager S.~C., Somerville R.~S., Gibson B.~K., 2010, MNRAS, 402,
  173

\bibitem[{Barnes \& Hernquist(1996)}]{Barnes:1996b}
Barnes J.~E., Hernquist L., 1996, AJ v.471, 471, 115

\bibitem[{{Behroozi} {et~al}\mbox{.}(2013{\natexlab{a}}){Behroozi}, {Wechsler},
  {Wu}, {Busha}, {Klypin}, \& {Primack}}]{Behroozi:2013a}
{Behroozi} P.~S., {Wechsler} R.~H., {Wu} H.-Y., {Busha} M.~T., {Klypin} A.~A.,
  {Primack} J.~R., 2013{\natexlab{a}}, ApJ, 763, 18

\bibitem[{{Behroozi} {et~al}\mbox{.}(2013{\natexlab{b}}){Behroozi}, {Wechsler},
  {Wu}, {Busha}, {Klypin}, \& {Primack}}]{behroozi:13}
{Behroozi} P.~S., {Wechsler} R.~H., {Wu} H.-Y., {Busha} M.~T., {Klypin} A.~A.,
  {Primack} J.~R., 2013{\natexlab{b}}, \apj, 763, 18

\bibitem[{{Bell} {et~al}\mbox{.}(2007){Bell}, {Zheng}, {Papovich}, {Borch},
  {Wolf}, \& {Meisenheimer}}]{bell:07}
{Bell} E.~F., {Zheng} X.~Z., {Papovich} C., {Borch} A., {Wolf} C.,
  {Meisenheimer} K., 2007, ApJ, 663, 834

\bibitem[{Bezanson {et~al}\mbox{.}(2011)Bezanson, van Dokkum, Franx, Brammer,
  Brinchmann, Kriek, Labb{\'e}, Quadri, Rix, van~de Sande, Whitaker, \&
  Williams}]{Bezanson:2011a}
Bezanson R. {et~al.}, 2011, ApJL, 737, L31

\bibitem[{Bezanson {et~al}\mbox{.}(2009)Bezanson, van Dokkum, Tal, Marchesini,
  Kriek, Franx, \& Coppi}]{Bezanson:2009a}
Bezanson R., van Dokkum P.~G., Tal T., Marchesini D., Kriek M., Franx M., Coppi
  P., 2009, ApJ, 697, 1290

\bibitem[{Bournaud {et~al}\mbox{.}(2011)Bournaud, Chapon, Teyssier, Powell,
  Elmegreen, Elmegreen, Duc, Contini, Epinat, \& Shapiro}]{Bournaud:2011a}
Bournaud F. {et~al.}, 2011, ApJ, 730, 4

\bibitem[{Bower {et~al}\mbox{.}(2006)Bower, Benson, Malbon, Helly, Frenk,
  Baugh, Cole, \& Lacey}]{Bower:2006a}
Bower R.~G., Benson A.~J., Malbon R., Helly J.~C., Frenk C.~S., Baugh C.~M.,
  Cole S., Lacey C.~G., 2006, MNRAS, 370, 645

\bibitem[{Bruzual \& Charlot(2003)}]{Bruzual:2003b}
Bruzual G., Charlot S., 2003, MNRAS, 344, 1000

\bibitem[{Cenarro \& Trujillo(2009)}]{Cenarro:2009a}
Cenarro A.~J., Trujillo I., 2009, ApJL, 696, L43

\bibitem[{Chabrier(2003)}]{Chabrier:2003a}
Chabrier G., 2003, PASP, 115, 763

\bibitem[{Cheng {et~al}\mbox{.}(2011)Cheng, Faber, Simard, Graves, Lopez, Yan,
  \& Cooper}]{Cheng:2011a}
Cheng J.~Y., Faber S.~M., Simard L., Graves G.~J., Lopez E.~D., Yan R., Cooper
  M.~C., 2011, MNRAS, 412, 727

\bibitem[{{Ciotti} {et~al}\mbox{.}(2007){Ciotti}, {Lanzoni}, \&
  {Volonteri}}]{Ciotti:2007a}
{Ciotti} L., {Lanzoni} B., {Volonteri} M., 2007, ApJ, 658, 65

\bibitem[{{Cole} {et~al}\mbox{.}(2000){Cole}, {Lacey}, {Baugh}, \&
  {Frenk}}]{Cole:2000a}
{Cole} S., {Lacey} C.~G., {Baugh} C.~M., {Frenk} C.~S., 2000, MNRAS, 319, 168

\bibitem[{{Combes} {et~al}\mbox{.}(1990){Combes}, {Debbasch}, {Friedli}, \&
  {Pfenniger}}]{combes:90}
{Combes} F., {Debbasch} F., {Friedli} D., {Pfenniger} D., 1990, \aap, 233, 82

\bibitem[{{Conroy} \& {van Dokkum}(2012)}]{Conroy:2012a}
{Conroy} C., {van Dokkum} P.~G., 2012, ApJ, 760, 71

\bibitem[{Covington {et~al}\mbox{.}(2008)Covington, Dekel, Cox, Jonsson, \&
  Primack}]{Covington:2008b}
Covington M., Dekel A., Cox T.~J., Jonsson P., Primack J.~R., 2008, MNRAS, 384,
  94

\bibitem[{{Covington} {et~al}\mbox{.}(2011){Covington}, {Primack}, {Porter},
  {Croton}, {Somerville}, \& {Dekel}}]{Covington:2011a}
{Covington} M.~D., {Primack} J.~R., {Porter} L.~A., {Croton} D.~J.,
  {Somerville} R.~S., {Dekel} A., 2011, MNRAS, 1029

\bibitem[{{Cox}(2004)}]{Cox:2004b}
{Cox} T.~J., 2004, PhD thesis, {UC Santa Cruz}

\bibitem[{Cox {et~al}\mbox{.}(2006)Cox, Dutta, Matteo, Hernquist, Hopkins,
  Robertson, \& Springel}]{Cox:2006a}
Cox T.~J., Dutta S.~N., Matteo T.~D., Hernquist L., Hopkins P.~F., Robertson
  B., Springel V., 2006, ApJ, 650, 791

\bibitem[{Cox {et~al}\mbox{.}(2008)Cox, Jonsson, Somerville, Primack, \&
  Dekel}]{Cox:2008b}
Cox T.~J., Jonsson P., Somerville R.~S., Primack J.~R., Dekel A., 2008, MNRAS,
  384, 386

\bibitem[{Croton {et~al}\mbox{.}(2006)Croton, Springel, White, {De Lucia},
  Frenk, Gao, Jenkins, Kauffmann, Navarro, \& Yoshida}]{Croton:2006a}
Croton D.~J. {et~al.}, 2006, MNRAS, 365, 11

\bibitem[{De~Lucia {et~al}\mbox{.}(2011)De~Lucia, Fontanot, Wilman, \&
  Monaco}]{De-Lucia:2011a}
De~Lucia G., Fontanot F., Wilman D., Monaco P., 2011, MNRAS, 517

\bibitem[{{Debattista} {et~al}\mbox{.}(2004){Debattista}, {Carollo}, {Mayer},
  \& {Moore}}]{debattista:04}
{Debattista} V.~P., {Carollo} C.~M., {Mayer} L., {Moore} B., 2004, \apjl, 604,
  L93

\bibitem[{Dekel {et~al}\mbox{.}(2009)Dekel, Birnboim, Engel, Freundlich,
  Goerdt, Mumcuoglu, Neistein, Pichon, Teyssier, \& Zinger}]{Dekel:2009a}
Dekel A. {et~al.}, 2009, Nat, 457, 451

\bibitem[{Dekel \& Cox(2006)}]{Dekel:2006a}
Dekel A., Cox T.~J., 2006, MNRAS, 370, 1445

\bibitem[{Dekel {et~al}\mbox{.}(2013)Dekel, Zolotov, Tweed, Cacciato, Ceverino,
  \& Primack}]{Dekel:2013a}
Dekel A., Zolotov A., Tweed D., Cacciato M., Ceverino D., Primack J.~R., 2013,
  ArXiv, 3009

\bibitem[{Djorgovski \& Davis(1987)}]{Djorgovski:1987b}
Djorgovski S., Davis M., 1987, ApJ, 313, 59

\bibitem[{{Dressler} {et~al}\mbox{.}(1987){Dressler}, {Lynden-Bell},
  {Burstein}, {Davies}, {Faber}, {Terlevich}, \& {Wegner}}]{Dressler:1987a}
{Dressler} A., {Lynden-Bell} D., {Burstein} D., {Davies} R.~L., {Faber} S.~M.,
  {Terlevich} R., {Wegner} G., 1987, ApJ, 313, 42

\bibitem[{Dutton {et~al}\mbox{.}(2012)Dutton, Mendel, \& Simard}]{Dutton:2012b}
Dutton A.~A., Mendel J.~T., Simard L., 2012, MNRAS:L, 422, L33

\bibitem[{Efstathiou {et~al}\mbox{.}(1982)Efstathiou, Lake, \&
  Negroponte}]{Efstathiou:1982a}
Efstathiou G., Lake G., Negroponte J., 1982, MNRAS, 199, 1069

\bibitem[{Faber {et~al}\mbox{.}(1987)Faber, Dressler, Davies, Burstein, \&
  Lynden-Bell}]{Faber:1987a}
Faber S.~M., Dressler A., Davies R.~L., Burstein D., Lynden-Bell D., 1987, in
  Nearly Normal Galaxies. From the Planck Time to the Present, ed. S. M. Faber
  (New York, NY: Springer-Verlag), 175

\bibitem[{{Faber} {et~al}\mbox{.}(2007){Faber} {et~al.}}]{faber:07}
{Faber} S.~M., {et~al.}, 2007, \apj, 665, 265

\bibitem[{{Fontanot} {et~al}\mbox{.}(2009){Fontanot}, {De Lucia}, {Monaco},
  {Somerville}, \& {Santini}}]{Fontanot:2009a}
{Fontanot} F., {De Lucia} G., {Monaco} P., {Somerville} R.~S., {Santini} P.,
  2009, MNRAS, 397, 1776

\bibitem[{Forbes {et~al}\mbox{.}(1998)Forbes, Ponman, \& Brown}]{Forbes:1998a}
Forbes D.~A., Ponman T.~J., Brown R. J.~N., 1998, ApJ, 508, L43

\bibitem[{Gallazzi {et~al}\mbox{.}(2005)Gallazzi, Charlot, Brinchmann, White,
  \& Tremonti}]{Gallazzi:2005a}
Gallazzi A., Charlot S., Brinchmann J., White S. D.~M., Tremonti C.~A., 2005,
  MNRAS, 362, 41

\bibitem[{Graves \& Faber(2010)}]{Graves:2010a}
Graves G.~J., Faber S.~M., 2010, ApJ, 717, 803

\bibitem[{Graves {et~al}\mbox{.}(2009{\natexlab{a}})Graves, Faber, \&
  Schiavon}]{Graves:2009c}
Graves G.~J., Faber S.~M., Schiavon R.~P., 2009{\natexlab{a}}, ApJ, 693, 486

\bibitem[{Graves {et~al}\mbox{.}(2009{\natexlab{b}})Graves, Faber, \&
  Schiavon}]{Graves:2009b}
Graves G.~J., Faber S.~M., Schiavon R.~P., 2009{\natexlab{b}}, ApJ, 698, 1590

\bibitem[{Graves {et~al}\mbox{.}(2010)Graves, Faber, \&
  Schiavon}]{Graves:2010b}
Graves G.~J., Faber S.~M., Schiavon R.~P., 2010, ApJ, 721, 278

\bibitem[{{Greene} {et~al}\mbox{.}(2012){Greene}, {Murphy}, {Comerford},
  {Gebhardt}, \& {Adams}}]{Greene:2012}
{Greene} J.~E., {Murphy} J.~D., {Comerford} J.~M., {Gebhardt} K., {Adams}
  J.~J., 2012, \apj, 750, 32

\bibitem[{Guo {et~al}\mbox{.}(2011)Guo, White, Boylan-Kolchin, De~Lucia,
  Kauffmann, Lemson, Li, Springel, \& Weinmann}]{Guo:2011a}
Guo Q. {et~al.}, 2011, MNRAS, 164

\bibitem[{{Hilz} {et~al}\mbox{.}(2013){Hilz}, {Naab}, \&
  {Ostriker}}]{Hilz:2013}
{Hilz} M., {Naab} T., {Ostriker} J.~P., 2013, \mnras, 429, 2924

\bibitem[{{Hohl}(1971)}]{hohl:71}
{Hohl} F., 1971, \apj, 168, 343

\bibitem[{Hopkins \& Beacom(2008)}]{Hopkins:2008a}
Hopkins A.~M., Beacom J.~F., 2008, ApJ, 682, 1486

\bibitem[{Hopkins {et~al}\mbox{.}(2010)Hopkins, Bundy, Hernquist, Wuyts, \&
  Cox}]{Hopkins:2010b}
Hopkins P.~F., Bundy K., Hernquist L., Wuyts S., Cox T.~J., 2010, MNRAS, 401,
  1099

\bibitem[{Johansson {et~al}\mbox{.}(2009)Johansson, Naab, \&
  Burkert}]{Johansson:2009a}
Johansson P.~H., Naab T., Burkert A., 2009, ApJ, 690, 802

\bibitem[{{Johansson} {et~al}\mbox{.}(2012){Johansson}, {Naab}, \&
  {Ostriker}}]{Johansson:2012b}
{Johansson} P.~H., {Naab} T., {Ostriker} J.~P., 2012, \apj, 754, 115

\bibitem[{Jones {et~al}\mbox{.}(2009)Jones, Read, Saunders, Colless, Jarrett,
  Parker, Fairall, Mauch, \& Sadler}]{Jones:2009a}
Jones D.~H. {et~al.}, 2009, MNRAS, 399, 683

\bibitem[{Jones {et~al}\mbox{.}(2004)Jones, Saunders, Colless, Read, Parker,
  Watson, Campbell, Burkey, Mauch, Moore, Hartley, Cass, James, Russell,
  Fiegert, Dawe, Huchra, Jarrett, Lahav, Lucey, Mamon, Proust, Sadler, \&
  Wakamatsu}]{Jones:2004a}
Jones D.~H. {et~al.}, 2004, MNRAS, 355, 747

\bibitem[{J{\o}rgensen {et~al}\mbox{.}(2006)J{\o}rgensen, Chiboucas, Flint,
  Bergmann, Barr, \& Davies}]{Jorgensen:2006a}
J{\o}rgensen I., Chiboucas K., Flint K., Bergmann M., Barr J., Davies R., 2006,
  ApJ, 639, L9

\bibitem[{J{\o}rgensen {et~al}\mbox{.}(1996)J{\o}rgensen, Franx, \&
  Kjaergaard}]{Jorgensen:1996b}
J{\o}rgensen I., Franx M., Kjaergaard P., 1996, MNRAS, 280, 167

\bibitem[{{Kauffmann} {et~al}\mbox{.}(2003){Kauffmann}, {Heckman}, {White},
  {Charlot}, {Tremonti}, {Peng}, {Seibert}, {Brinkmann}, {Nichol}, {SubbaRao},
  \& {York}}]{Kauffmann:2003a}
{Kauffmann} G. {et~al.}, 2003, MNRAS, 341, 54

\bibitem[{Khochfar \& Burkert(2003)}]{Khochfar:2003a}
Khochfar S., Burkert A., 2003, ApJ, 597, L117

\bibitem[{{Khochfar} \& {Silk}(2006)}]{khochfar:06}
{Khochfar} S., {Silk} J., 2006, ApJL, 648, L21

\bibitem[{Kirby {et~al}\mbox{.}(2010)Kirby, Guhathakurta, Simon, Geha, Rockosi,
  Sneden, Cohen, Sohn, Majewski, \& Siegel}]{Kirby:2010a}
Kirby E.~N. {et~al.}, 2010, The Astrophysical Journal Supplement, 191, 352

\bibitem[{{Klypin} {et~al}\mbox{.}(2011){Klypin}, {Trujillo-Gomez}, \&
  {Primack}}]{Klypin:2011a}
{Klypin} A.~A., {Trujillo-Gomez} S., {Primack} J., 2011, ApJ, 740, 102

\bibitem[{{Komatsu} {et~al}\mbox{.}(2009){Komatsu}, {Dunkley}, {Nolta},
  {Bennett}, {Gold}, {Hinshaw}, {Jarosik}, {Larson}, {Limon}, {Page},
  {Spergel}, {Halpern}, {Hill}, {Kogut}, {Meyer}, {Tucker}, {Weiland},
  {Wollack}, \& {Wright}}]{Komatsu:2009a}
{Komatsu} E. {et~al.}, 2009, ApJS, 180, 330

\bibitem[{{Komatsu} {et~al}\mbox{.}(2011){Komatsu}, {Smith}, {Dunkley},
  {Bennett}, {Gold}, {Hinshaw}, {Jarosik}, {Larson}, {Nolta}, {Page},
  {Spergel}, {Halpern}, {Hill}, {Kogut}, {Limon}, {Meyer}, {Odegard}, {Tucker},
  {Weiland}, {Wollack}, \& {Wright}}]{Komatsu:2011a}
{Komatsu} E. {et~al.}, 2011, ApJS, 192, 18

\bibitem[{{Laporte} {et~al}\mbox{.}(2013){Laporte}, {White}, {Naab}, \&
  {Gao}}]{Laporte:2013}
{Laporte} C.~F.~P., {White} S.~D.~M., {Naab} T., {Gao} L., 2013, \mnras, 435,
  901

\bibitem[{Magoulas {et~al}\mbox{.}(2012)Magoulas, Springob, Colless, Jones,
  Campbell, Lucey, Mould, Jarrett, Merson, \& Brough}]{Magoulas:2012b}
Magoulas C. {et~al.}, 2012, MNRAS, 427, 245

\bibitem[{{Mihos} \& {Hernquist}(1994)}]{Mihos:1994a}
{Mihos} J.~C., {Hernquist} L., 1994, ApJ, 437, L47

\bibitem[{{Naab} {et~al}\mbox{.}(2009){Naab}, {Johansson}, \&
  {Ostriker}}]{Naab:2009a}
{Naab} T., {Johansson} P.~H., {Ostriker} J.~P., 2009, ApJL, 699, L178

\bibitem[{Naab {et~al}\mbox{.}(2006)Naab, Khochfar, \& Burkert}]{Naab:2006b}
Naab T., Khochfar S., Burkert A., 2006, ApJ, 636, L81

\bibitem[{Nelan {et~al}\mbox{.}(2005)Nelan, Smith, Hudson, Wegner, Lucey,
  Moore, Quinney, \& Suntzeff}]{Nelan:2005a}
Nelan J.~E., Smith R.~J., Hudson M.~J., Wegner G.~A., Lucey J.~R., Moore S.
  A.~W., Quinney S.~J., Suntzeff N.~B., 2005, ApJ, 632, 137

\bibitem[{Newman {et~al}\mbox{.}(2012)Newman, Ellis, Bundy, \&
  Treu}]{Newman:2012b}
Newman A.~B., Ellis R.~S., Bundy K., Treu T., 2012, ApJ, 746, 162

\bibitem[{Nipoti {et~al}\mbox{.}(2012)Nipoti, Treu, Leauthaud, Bundy, Newman,
  \& Auger}]{Nipoti:2012a}
Nipoti C., Treu T., Leauthaud A., Bundy K., Newman A.~B., Auger M.~W., 2012,
  MNRAS, 422, 1714

\bibitem[{Oser {et~al}\mbox{.}(2012)Oser, Naab, Ostriker, \&
  Johansson}]{Oser:2012a}
Oser L., Naab T., Ostriker J.~P., Johansson P.~H., 2012, ApJ, 744, 63

\bibitem[{{Ostriker} \& {Peebles}(1973)}]{OP:73}
{Ostriker} J.~P., {Peebles} P.~J.~E., 1973, \apj, 186, 467

\bibitem[{{Parry} {et~al}\mbox{.}(2009){Parry}, {Eke}, \& {Frenk}}]{parry:09}
{Parry} O.~H., {Eke} V.~R., {Frenk} C.~S., 2009, \mnras, 396, 1972

\bibitem[{Porter {et~al}\mbox{.}(2014)Porter, Somerville, Primack, \&
  Johannson}]{Porter:2014a}
Porter L., Somerville R.~S., Primack J.~R., Johannson P., 2014, submitted

\bibitem[{Quilis \& Trujillo(2012)}]{Quilis:2012a}
Quilis V., Trujillo I., 2012, ApJL, 752, L19

\bibitem[{Shankar {et~al}\mbox{.}(2011)Shankar, Marulli, Bernardi, Mei, Meert,
  \& Vikram}]{Shankar:2011a}
Shankar F., Marulli F., Bernardi M., Mei S., Meert A., Vikram V., 2011, ArXiv,
  1105, 6043

\bibitem[{Smith {et~al}\mbox{.}(2007)Smith, Lucey, \& Hudson}]{Smith:2007a}
Smith R.~J., Lucey J.~R., Hudson M.~J., 2007, MNRAS, 381, 1035

\bibitem[{Somerville {et~al}\mbox{.}(2012)Somerville, Gilmore, Primack, \&
  Dom{\'\i}nguez}]{Somerville:2012a}
Somerville R.~S., Gilmore R.~C., Primack J.~R., Dom{\'\i}nguez A., 2012, MNRAS,
  2820

\bibitem[{Somerville {et~al}\mbox{.}(2008)Somerville, Hopkins, Cox, Robertson,
  \& Hernquist}]{Somerville:2008a}
Somerville R.~S., Hopkins P.~F., Cox T.~J., Robertson B.~E., Hernquist L.,
  2008, MNRAS, 391, 481

\bibitem[{Spiniello {et~al}\mbox{.}(2012)Spiniello, Trager, Koopmans, \&
  Chen}]{Spiniello:2012a}
Spiniello C., Trager S.~C., Koopmans L. V.~E., Chen Y.~P., 2012, ApJL, 753, L32

\bibitem[{Springob {et~al}\mbox{.}(2012)Springob, Magoulas, Proctor, Colless,
  Jones, Kobayashi, Campbell, Lucey, \& Mould}]{Springob:2012a}
Springob C.~M. {et~al.}, 2012, MNRAS, 420, 2773

\bibitem[{Strauss {et~al}\mbox{.}(2002)Strauss, Weinberg, Lupton, Narayanan,
  Annis, Bernardi, York, \& Zehavi}]{Strauss:2002a}
Strauss M.~A., Weinberg D.~H., Lupton R.~H., Narayanan V.~K., Annis J.,
  Bernardi M., York D.~G., Zehavi I., 2002, AJ, 124, 1810

\bibitem[{Terlevich \& Forbes(2002)}]{Terlevich:2002a}
Terlevich A.~I., Forbes D.~A., 2002, MNRAS, 330, 547

\bibitem[{{Thomas} {et~al}\mbox{.}(2010){Thomas}, {Maraston}, {Schawinski},
  {Sarzi}, \& {Silk}}]{Thomas:2010}
{Thomas} D., {Maraston} C., {Schawinski} K., {Sarzi} M., {Silk} J., 2010,
  \mnras, 404, 1775

\bibitem[{Toomre(1964)}]{Toomre:1964a}
Toomre A., 1964, ApJ, 139, 1217

\bibitem[{{Toomre}(1977)}]{Toomre:1977a}
{Toomre} A., 1977, in Evolution of Galaxies and Stellar Populations, p. 401

\bibitem[{{Toomre} \& {Toomre}(1972)}]{Toomre:1972a}
{Toomre} A., {Toomre} J., 1972, ApJ, 178, 623

\bibitem[{Trager \& Somerville(2009)}]{Trager:2009a}
Trager S.~C., Somerville R.~S., 2009, MNRAS, 395, 608

\bibitem[{Treu {et~al}\mbox{.}(2005{\natexlab{a}})Treu, Ellis, Liao, \& van
  Dokkum}]{Treu:2005a}
Treu T., Ellis R.~S., Liao T.~X., van Dokkum P.~G., 2005{\natexlab{a}}, ApJ,
  622, L5

\bibitem[{Treu {et~al}\mbox{.}(2005{\natexlab{b}})Treu, Ellis, Liao, van
  Dokkum, Tozzi, Coil, Newman, Cooper, \& Davis}]{Treu:2005b}
Treu T. {et~al.}, 2005{\natexlab{b}}, ApJ, 633, 174

\bibitem[{Trujillo {et~al}\mbox{.}(2011)Trujillo, Ferreras, \& de~la
  Rosa}]{Trujillo:2011a}
Trujillo I., Ferreras I., de~la Rosa I.~G., 2011, MNRAS, 415, 3903

\bibitem[{{Trujillo-Gomez} {et~al}\mbox{.}(2011){Trujillo-Gomez}, {Klypin},
  {Primack}, \& {Romanowsky}}]{Trujillo-Gomez:2011a}
{Trujillo-Gomez} S., {Klypin} A., {Primack} J., {Romanowsky} A.~J., 2011, ApJL,
  742, 16

\bibitem[{van~der Wel {et~al}\mbox{.}(2004)van~der Wel, Franx, van Dokkum, \&
  Rix}]{Wel:2004a}
van~der Wel A., Franx M., van Dokkum P.~G., Rix H.-W., 2004, ApJ, 601, L5

\bibitem[{van Dokkum \& van~der Marel(2007)}]{Dokkum:2007a}
van Dokkum P.~G., van~der Marel R.~P., 2007, ApJ, 655, 30

\bibitem[{Wechsler {et~al}\mbox{.}(2002)Wechsler, Bullock, Primack, Kravtsov,
  \& Dekel}]{Wechsler:2002b}
Wechsler R.~H., Bullock J.~S., Primack J.~R., Kravtsov A.~V., Dekel A., 2002,
  ApJ, 568, 52

\bibitem[{Woo {et~al}\mbox{.}(2008)Woo, Courteau, \& Dekel}]{Woo:2008a}
Woo J., Courteau S., Dekel A., 2008, Monthly Notices of the Royal Astronomical
  Society, 390, 1453

\bibitem[{Worthey {et~al}\mbox{.}(1994)Worthey, Faber, Gonzalez, \&
  Burstein}]{Worthey:1994a}
Worthey G., Faber S.~M., Gonzalez J.~J., Burstein D., 1994, ApJS, 94, 687

\bibitem[{Worthey \& Ottaviani(1997)}]{Worthey:1997a}
Worthey G., Ottaviani D.~L., 1997, ApJS, 111, 377

\bibitem[{Wuyts {et~al}\mbox{.}(2010)Wuyts, Cox, Hayward, Franx, Hernquist,
  Hopkins, Jonsson, \& van Dokkum}]{Wuyts:2010b}
Wuyts S., Cox T.~J., Hayward C.~C., Franx M., Hernquist L., Hopkins P.~F.,
  Jonsson P., van Dokkum P.~G., 2010, ApJ, 722, 1666

\bibitem[{{Yates} \& {Kauffmann}(2014)}]{Yates:14}
{Yates} R.~M., {Kauffmann} G., 2014, \mnras, 439, 3817

\bibitem[{York {et~al}\mbox{.}(2000)York, Adelman, Anderson, Anderson, Yanny,
  \& Yasuda}]{York:2000a}
York D.~G., Adelman J., Anderson J.~E., Anderson S.~F., Yanny B., Yasuda N.,
  2000, AJ, 120, 1579

\end{thebibliography}

\end{document}